\documentclass[fleqn,usenatbib]{mnras}

\usepackage{newtxtext,newtxmath}

\usepackage[english]{babel}
\usepackage[utf8]{inputenc}
\usepackage{multirow}

\usepackage[T1]{fontenc}
\usepackage{ae,aecompl}

\usepackage{graphicx}	
\usepackage{amsmath}	
\usepackage{amssymb}	
\usepackage{subfigure}

\title[The first stellar occultations by Phoebe]{The first observed stellar occultations by the irregular satellite \textcolor{red}{(Saturn IX)} Phoebe and improved rotational period}

\author[Gomes-Júnior et al.]{A. R. Gomes-Júnior$^{1,2,3,}$\thanks{E-mail: altairgomesjr@gmail.com, altair.gomes@unesp.br, altair.gomes@linea.gov.br},
M. Assafin$^{2,3,\ddag}$, 
F. Braga-Ribas$^{4,5,3,6}$, 
G. Benedetti-Rossi$^{6,5,3}$,\newauthor 
B. Morgado$^{5,3}$, 
J. I. B. Camargo$^{5,3}$, 
R. Vieira-Martins$^{5,3,2,\ddag}$, 
J. Desmars$^{7,8}$, 
B. Sicardy$^{6}$, \newauthor 
T. Barry$^{9}$,  
J. Campbell-White$^{10}$, 
E. Fernández-Lajús$^{11,12}$,  
D. Giles$^{9}$,  
W. Hanna$^{13}$, \newauthor 
T. Hayamizu$^{14}$, 
T. Hirose$^{14}$,  
A. De Horta$^{9}$,  
R. Horvat$^{9}$,  
K. Hosoi$^{14}$,  
E. Jehin$^{15}$,  
S. Kerr$^{16,17}$, \newauthor 
D. I. Machado$^{18,19}$,   
L. A. Mammana$^{11,20}$,  
D. Maybour$^{9}$, 
M. Owada$^{14}$, 
S. Rahvar$^{21}$ \newauthor 
C. Snodgrass$^{22}$ 
\\
$^{1}$UNESP - São Paulo State University, Grupo de Dinâmica Orbital e Planetologia, CEP 12516-410, Guaratinguetá, SP, Brazil\\
$^{2}$Observatório do Valongo/UFRJ, Ladeira Pedro Ant\^onio 43, CEP 20.080-090 Rio de Janeiro - RJ, Brazil\\
$^{3}$Laboratório Interinstitucional de e-Astronomia - LIneA, Rua Gal. José Cristino 77, Rio de Janeiro, RJ 20921-400, Brazil\\
$^{4}$Federal University of Technology - Paraná (UTFPR / DAFIS), Rua Sete de Setembro, 3165, CEP 80230-901, Curitiba, PR, Brazil\\
$^{5}$Observatório Nacional/MCTI, R. General José Cristino 77, CEP 20921-400 Rio de Janeiro - RJ, Brazil\\
$^{6}$LESIA, Observatoire de Paris – Section Meudon, 5 Place Jules Janssen – 92195 Meudon Cedex\\
$^{7}$Institut Poytechnique des Sciences Avanc\'ees IPSA, 63 boulevard de Brandebourg, F-94200 Ivry-sur-Seine, France\\
$^{8}$Institut de M\'ecanique C\'eleste et de Calcul des \'Eph\'em\'erides, IMCCE, Observatoire de Paris, PSL Research University, CNRS, Sorbonne Universités,\\UPMC Univ Paris 06, Univ. Lille, 77 Av. Denfert-Rochereau, F-75014 Paris, France;\\
$^{9}$Penrith Observatory, Western Sydney University, School of Computing, Engineering and Mathematics, Kingswood, NSW, AUS\\
$^{10}$SUPA, School of Science and Engineering, University of Dundee, Nethergate, Dundee DD1 4HN, U.K.\\
$^{11}$Facultad de Ciencias Astronómicas y Geofísicas - Universidad Nacional de La Plata, Paseo del Bosque S/N - 1900, La Plata, Argentina\\
$^{12}$Instituto de Astrof\'isica de La Plata (CCT La Plata - CONICET/UNLP), Argentina\\
$^{13}$Royal Astronomical Society of New Zealand, Occultation Section; International Occultation Timing Association (IOTA), Columbia Falls, MT 59912, USA\\
$^{14}$Japan Occultation Information Network (JOIN), Japan\\
$^{15}$Space sciences, Technologies \& Astrophysics Research (STAR) Institute, Université de Liège, 4000 Liège, Belgium\\
$^{16}$Astronomical Association of Queensland, QLD, Australia\\
$^{17}$Occultation Section of the Royal Astronomical Society of New Zealand (RASNZ), Wellington, New Zealand\\
$^{18}$Polo Astronômico Casimiro Montenegro Filho/FPTI-BR, Avenida Tancredo Neves 6731, CEP 85867-900, Foz do Iguaçu, PR, Brazil.\\
$^{19}$Universidade Estadual do Oeste do Paraná, Avenida Tarquínio Joslin dos Santos, 1300, CEP 85870-650, Foz do Iguaçu, PR, Brazil.\\
$^{20}$Complejo Astronómico El Leoncito (CASLEO), Av. España 1512 Sur, J5402DSP - San Juan – Argentina\\
$^{21}$Department of Physics, Sharif University of Technology, P.O. Box 11155-9161 Tehran, Iran\\
$^{22}$Institute for Astronomy, University of Edinburgh, Royal Observatory, Edinburgh EH9 3HJ, UK\\
$^\ddag$Affiliated researcher at Observatoire de Paris/IMCCE, 77 Avenue Denfert Rochereau 75014 Paris, France
}

\date{Accepted XXX. Received YYY; in original form ZZZ}

\pubyear{2019}

\begin{document}
\label{firstpage}
\pagerange{\pageref{firstpage}--\pageref{lastpage}}
\maketitle

\begin{abstract}
We report six stellar occultations by \textcolor{red}{(Saturn IX)} Phoebe, an irregular satellite of Saturn, obtained between mid-2017 and mid-2019. The 2017 July 06 event is the first stellar occultation by an irregular satellite ever observed. The occultation chords were compared to a 3D shape model of the satellite obtained from \textit{Cassini} observations. The rotation period available in the literature led to a sub-observer point at the moment of the observed occultations where the chords could not fit the 3D model. A procedure was developed to identify the correct sub-observer longitude. It allowed us to obtain \textcolor{red}{the} rotation period with improved precision \textcolor{red}{over currently known value from literature}. We show that the difference between the observed and the predicted sub-observer longitude suggests two possible solutions for \textcolor{red}{the} rotation period. By comparing these values with recently observed rotational light curves and single-chord stellar occultations, we can identify the best solution for Phoebe's rotational period as \textcolor{red}{$9.27365 \pm 0.00002$ h}. From the stellar occultations, we also obtained 6 geocentric astrometric positions in the ICRS as realised by the Gaia-DR2 with uncertainties at the 1-mas level.

\end{abstract}




\section{Introduction}
\label{Sec:Phoebe-intro}

Phoebe is the first irregular satellite to be discovered, in 1898, by William Henry Pickering \citep{Pickering1899}. It is also the first object to be identified with a retrograde orbit by \cite{Ross1905}. Until the year 2000, it was the only known Saturnian irregular satellite \citep{Gladman2001}.

Phoebe is the only irregular satellite to have been visited by a spacecraft. The visit was made by the \textit{Cassini-Huygens} Spacecraft\footnote{NASA/ESA/ASI mission to explore the Saturnian system. Website: \url{http://sci.esa.int/cassini-huygens/}} in 11 June 2004 \citep{Porco2005}, which observations resolved the shape of the object. Unfortunately, since it was a quick flyby, not all regions of Phoebe were observed. The maximum resolution of \textit{Cassini} observations was 13 m px$^{-1}$, however, as can be seen in \cite{Porco2005}, the resolution varies a lot depending on the latitude and longitude of Phoebe. In particular, the region close to the North Pole (latitudes higher than $+60\degr$) was always in the dark, and it was not imaged. An important point is that, during 2017-2019, the north region is visible from Earth.

From \textit{Cassini} observations, \cite{Thomas2010} determined the semi-axis of Phoebe as a=$109.4\pm1.4$ km, b=$108.5\pm0.6$ km and c=$101.8\pm0.3$ km. Considering the error bars of $a$ and $b$, Phoebe can be considered as an oblate spheroid.

Due to its orbital characteristics, it is believed Phoebe has been captured by Saturn during the evolution of the Solar System. This assumption is supported by \textit{Cassini} observation showing that Phoebe must have a composition different from other Saturnian satellites \citep{Johnson2005}. \cite{Clark2005} concluded that Phoebe has a surface that is covered by material of cometary or outer Solar System origin. 

The technique of stellar occultation, which can achieve kilometre accuracy \citep[e.g.][]{BragaRibas2014}, can be used to constrain the shape of Phoebe in the regions that were not observed by \textit{Cassini}. Since Phoebe's pole coordinates and rotational period are known, it is possible to associate the occultation chords with a specific latitude and longitude on Phoebe's surface.

In this regard, \cite{GomesJunior2016} predicted stellar occultations by Phoebe up to 2020. They showed that Saturn was crossing a region on the sky that had the Galactic Plane as background by the end-2017 and 2018, increasing the number of predicted events by a factor of 10. To have good predictions, the ephemeris of Phoebe was improved by using all observations available.

The first observed stellar occultation by Phoebe occurred in 2017 July 06. This occultation was observed in Japan by two sites. In 2018, other four occultations were observed, in South America (June 19) and Australia (June 26, July 03 and August 13). The sixth one was observed in 2019 June 07 in South America. Only the first occultation was multi-chord.

The observations details are presented in \autoref{Sec:Predic} and the occultation light curves are described in \autoref{Sec:Curves}. In \autoref{Sec:Phoebe-2017}, we show the reduction process of the two chords of the 2017 July 06 event comparing them with a 3D shape model. In \autoref{Sec:Phoebe-rotation} we discuss the improvement of the rotation model of Phoebe from the occultations results. In \autoref{Sec:Phoebe-2018} we discuss over the 2018 and 2019 events and how they could help to improve Phoebe's rotation period. In \autoref{Sec:Phoebe-light_curve} we compare our solutions to a rotational light curve of Phoebe. The results are presented in \autoref{Sec:Phoebe-results}. The conclusion and final remarks are set in \autoref{Sec:Phoebe-discussao}.

\section{Predictions and Observations}
\label{Sec:Predic}

The predictions for all the events were first made by \cite{GomesJunior2016} with UCAC4 stars \citep{Zacharias2013}. We then updated the star positions with newer catalogues as the occultation epoch approached. The first occultation by Phoebe was observed in 2017 July 06, which star position was updated using Gaia-DR1. Since the star was also a Tycho-2 star (TYC 6247-505-1), we benefited of the Tycho-Gaia Astrometric Solution \citep[TGAS,][]{Michalik2015}, with proper motions and parallax, deriving a better position than most of the Gaia-DR1 stars.

In 2018 April, Gaia-DR2 was published \citep{Brown2018}, so the next candidate stars' positions were updated allowing accurate predictions and the observation of five more occultations. For the fitting procedure, the Gaia-DR2 star positions were used for all six events. \autoref{Tab:Occ-stars} shows the geocentric ICRS coordinates of the stars at the occultation epoch, corrected from proper motion, parallax and radial velocity. The catalogued G magnitude of the stars are presented, which can be compared to Phoebe's V=16.6.

\begin{table*}
\caption{Occulted star parameters for each event. \label{Tab:Occ-stars}}
\begin{centering}
\begin{tabular}{cccccc}
\hline  \hline
Occ. Date & Gaia DR2 star designation & G Mag & Right Ascension ($\alpha$)* & Declination ($\delta$)* & Diam.** (km) \tabularnewline
\hline
2017 July 06 & 4117746607441803776 & 10.3 & 17$^{h}$ 31$^{m}$ 03\fs03947 $\pm$ 0.2 mas & -22\degr 00\arcmin 58\farcs0938 $\pm$ 0.2 mas & 1.18 \tabularnewline
2018 June 19 & 4090124156586307072 & 14.9 & 18$^{h}$ 26$^{m}$ 16\fs40061 $\pm$ 0.4 mas & -22\degr 24\arcmin 00\farcs4378 $\pm$ 0.3 mas & 0.15 \tabularnewline
2018 June 26 & 4089759672845036416 & 14.4 &  18$^{h}$ 24$^{m}$ 01\fs50061 $\pm$ 0.2 mas & -22\degr 26\arcmin 12\farcs1804 $\pm$ 0.2 mas & 0.19 \tabularnewline
2018 July 03 & 4089802618196107776 & 15.6 & 18$^{h}$ 22$^{m}$ 00\fs25777 $\pm$ 0.3 mas & -22\degr 28\arcmin 10\farcs7057 $\pm$ 0.3 mas & 0.24 \tabularnewline
2018 Aug 13 & 4066666389602814208 & 17.5 & 18$^{h}$ 12$^{m}$ 22\fs37073 $\pm$ 0.6 mas & -22\degr 38\arcmin 44\farcs2087 $\pm$ 0.6 mas & 0.14 \tabularnewline
2019 June 07 & 6772694935064300416 & 15.3 & 19$^{h}$ 21$^{m}$ 18\fs63201 $\pm$ 0.3 mas & -21\degr 44\arcmin 25\farcs3924 $\pm$ 0.3 mas & 0.10 \tabularnewline
\hline
\end{tabular}
\par
\end{centering}
*Gaia-DR2 star position at the time of occultation.
**Apparent star diameter at the distance of Phoebe.
\end{table*}

The positions of Phoebe were determined from the ephemeris published by \cite{GomesJunior2016}, PH15, and the planetary ephemeris JPL DE438. PH15 is an updated version of PH12 published by \cite{Desmars2013}.

\begin{figure*}
\centering
\subfigure[][2017 July 06]{\includegraphics[width=0.46\textwidth]{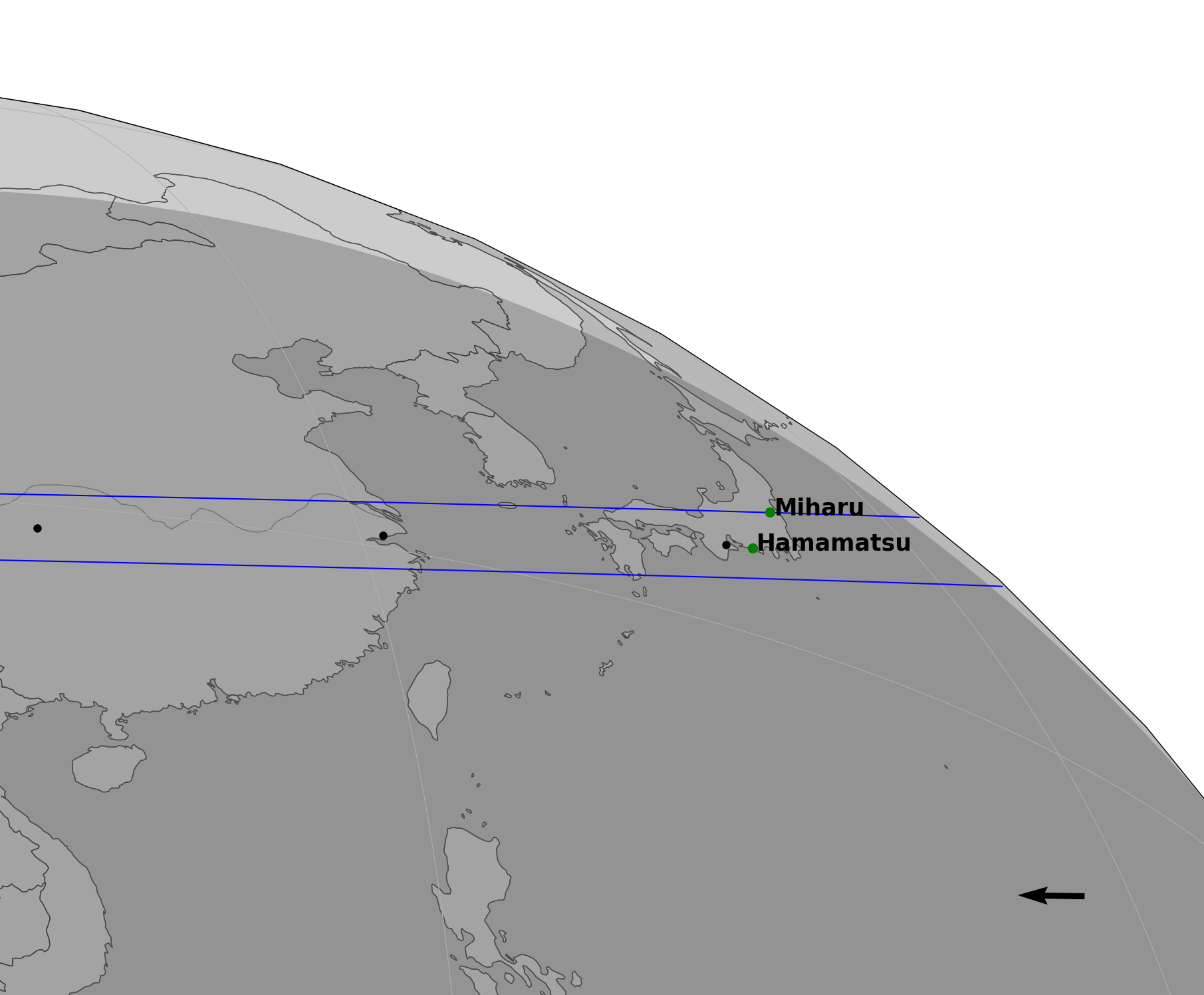}}
\subfigure[][2018 June 19]{\includegraphics[width=0.46\textwidth]{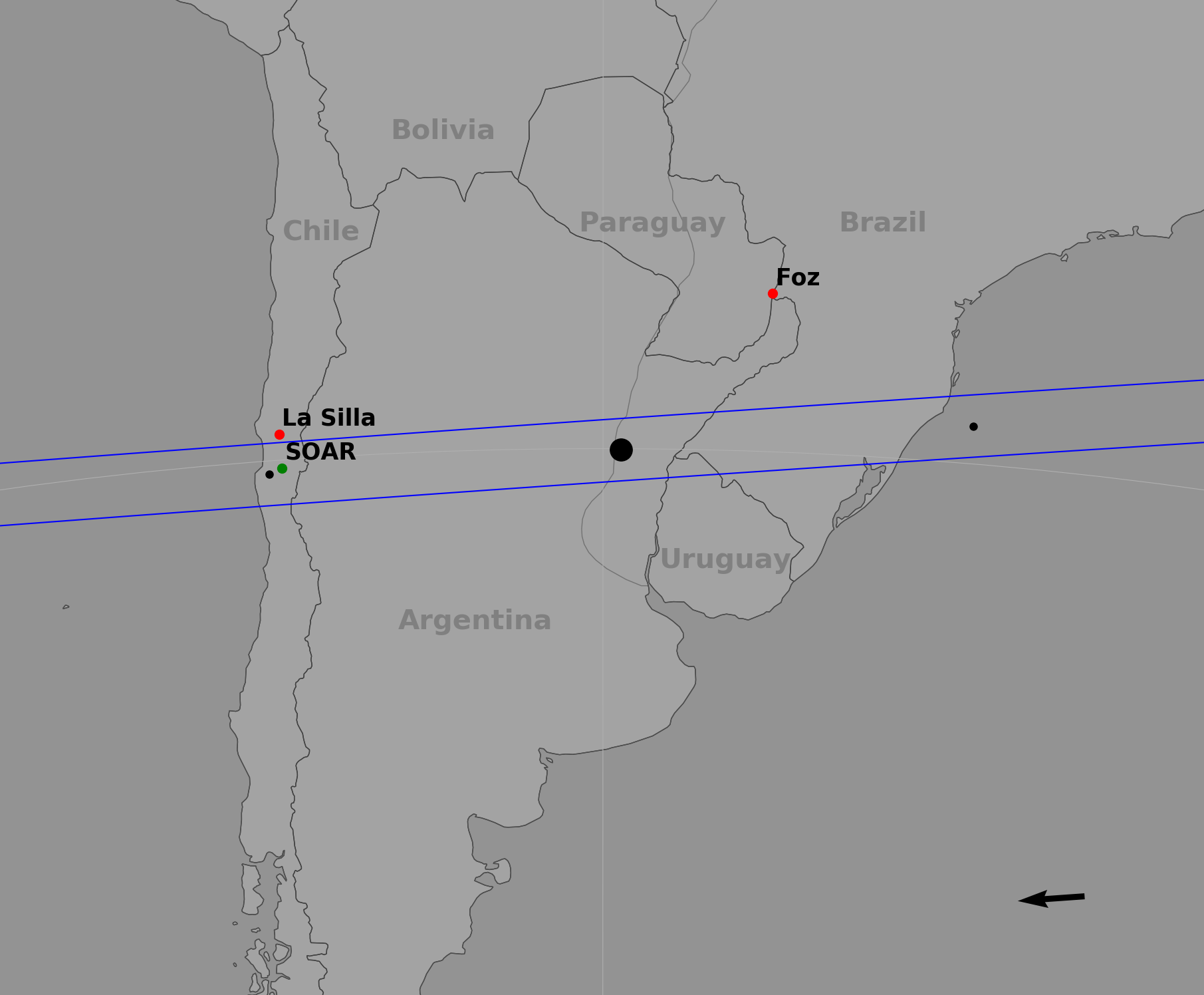}}
\subfigure[][2018 June 26]{\includegraphics[width=0.46\textwidth]{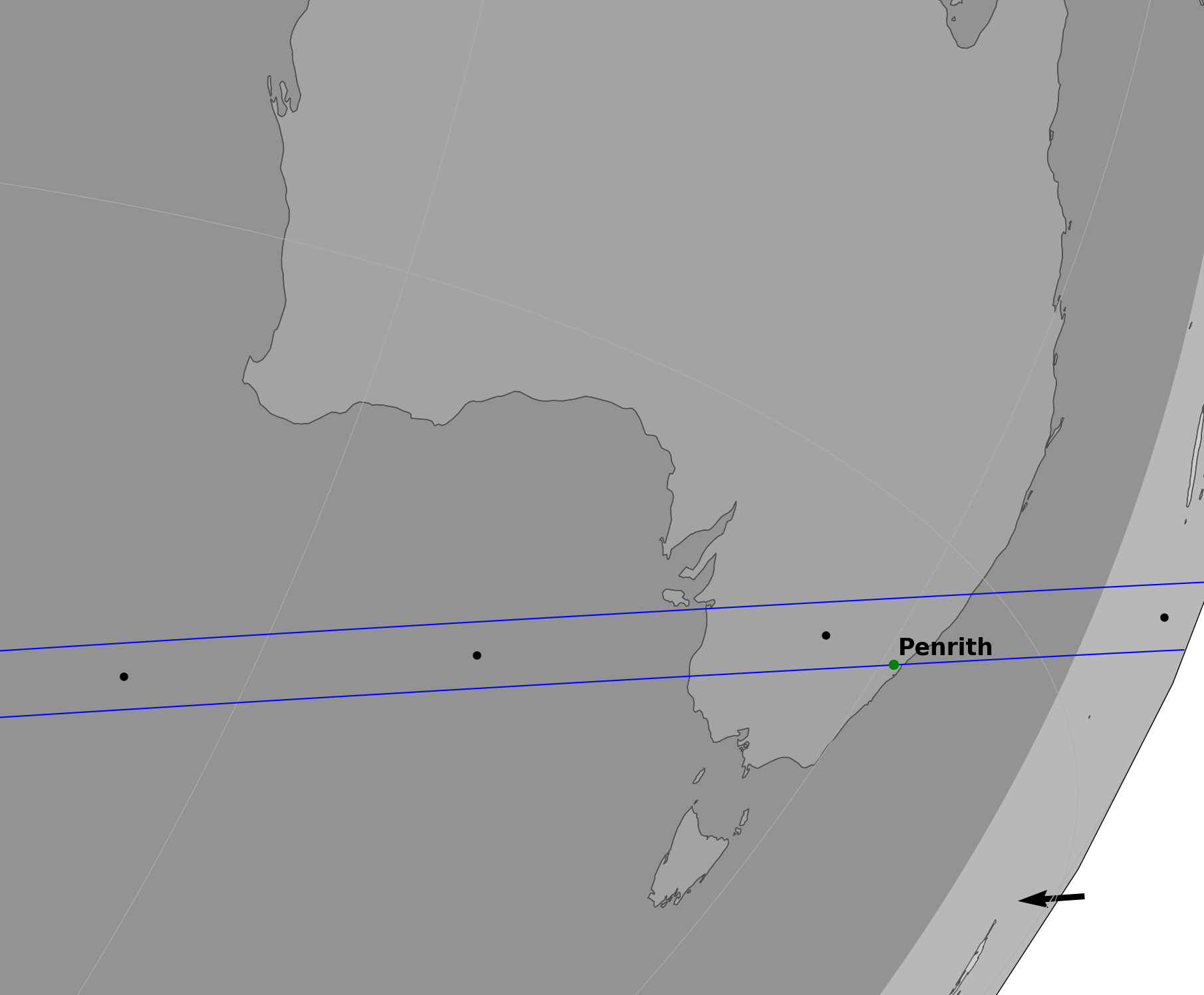}}
\subfigure[][2018 July 03]{\includegraphics[width=0.46\textwidth]{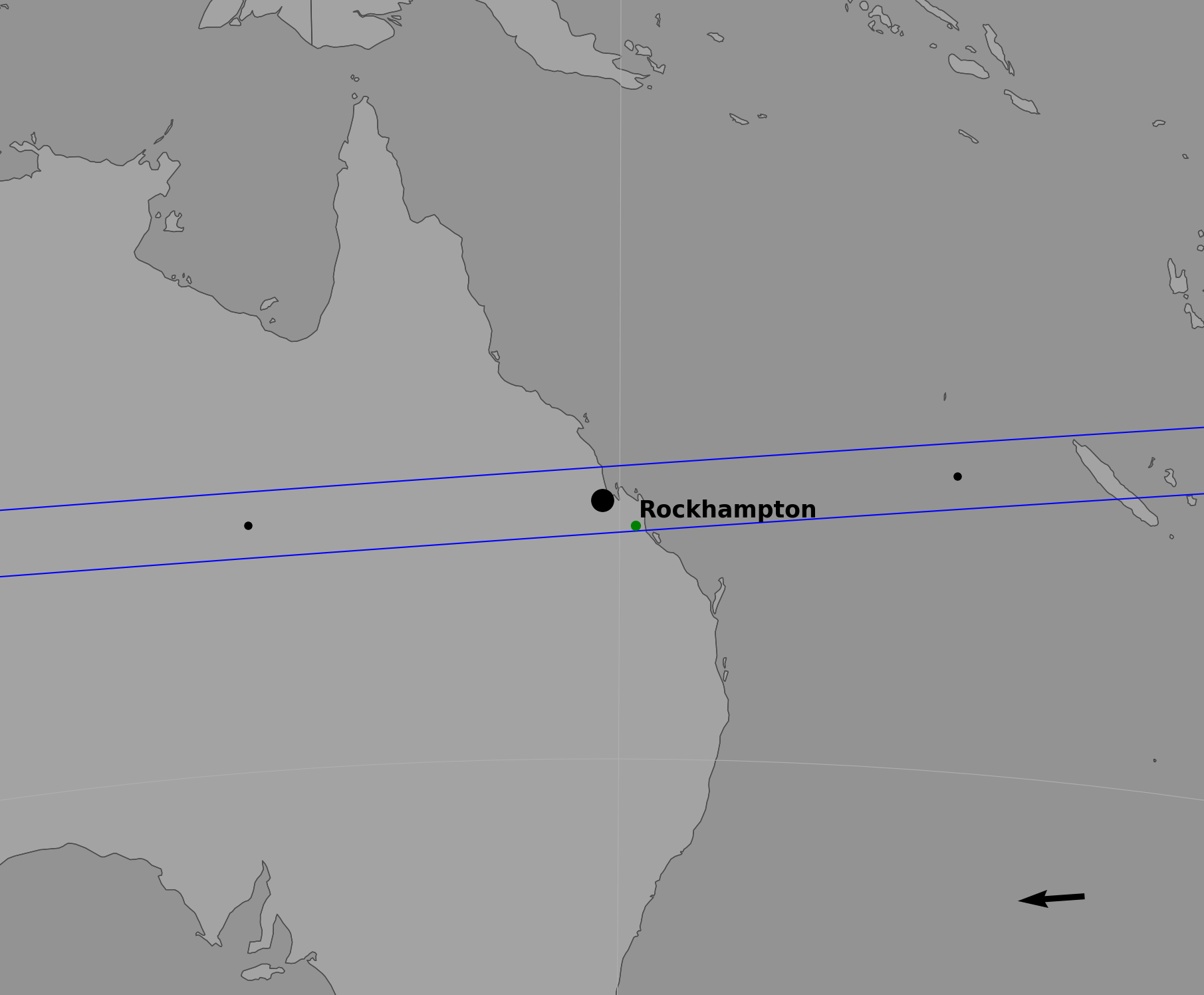}}
\subfigure[][2018 August 13]{\includegraphics[width=0.46\textwidth]{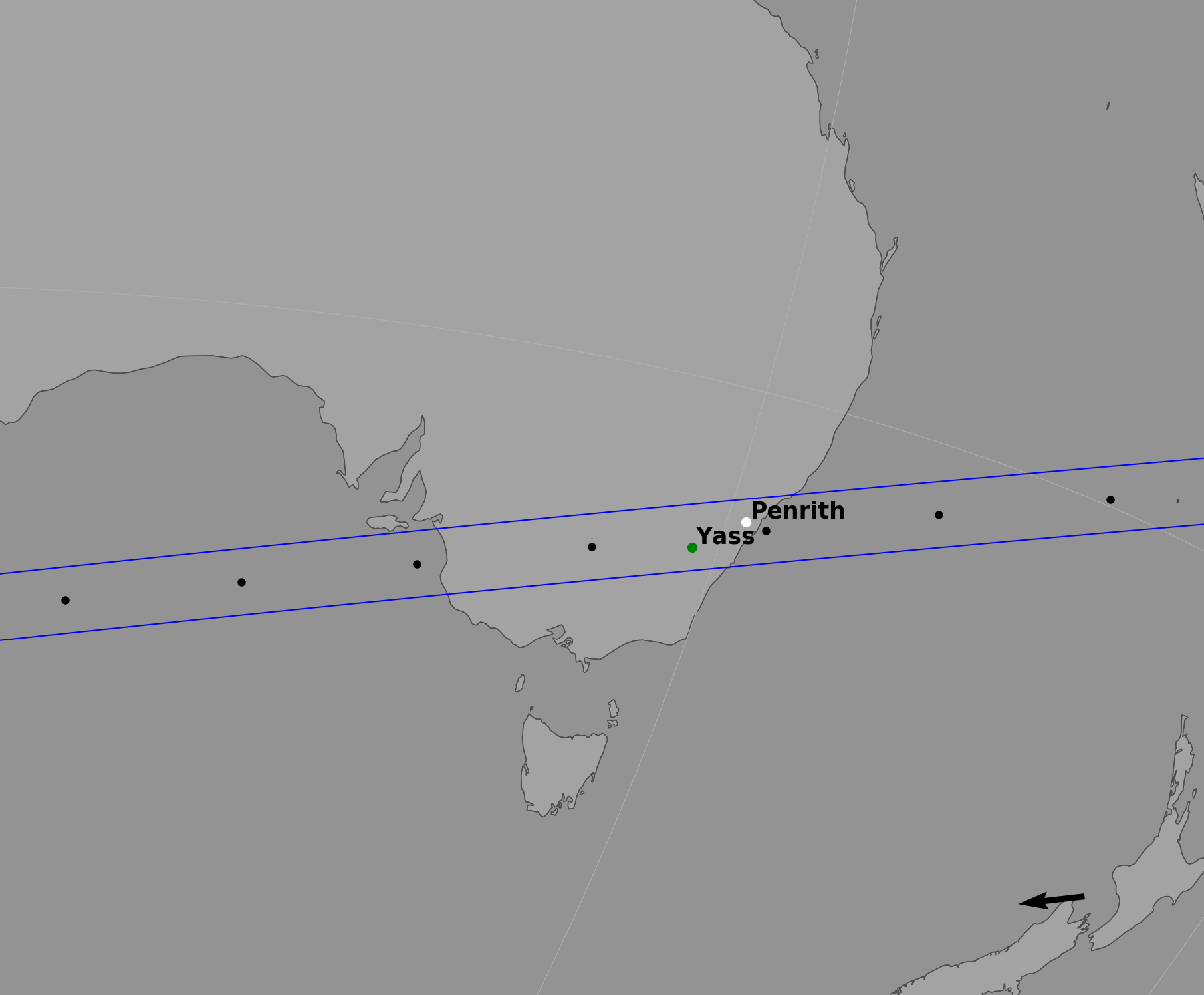}}
\subfigure[][2019 June 07]{\includegraphics[width=0.46\textwidth]{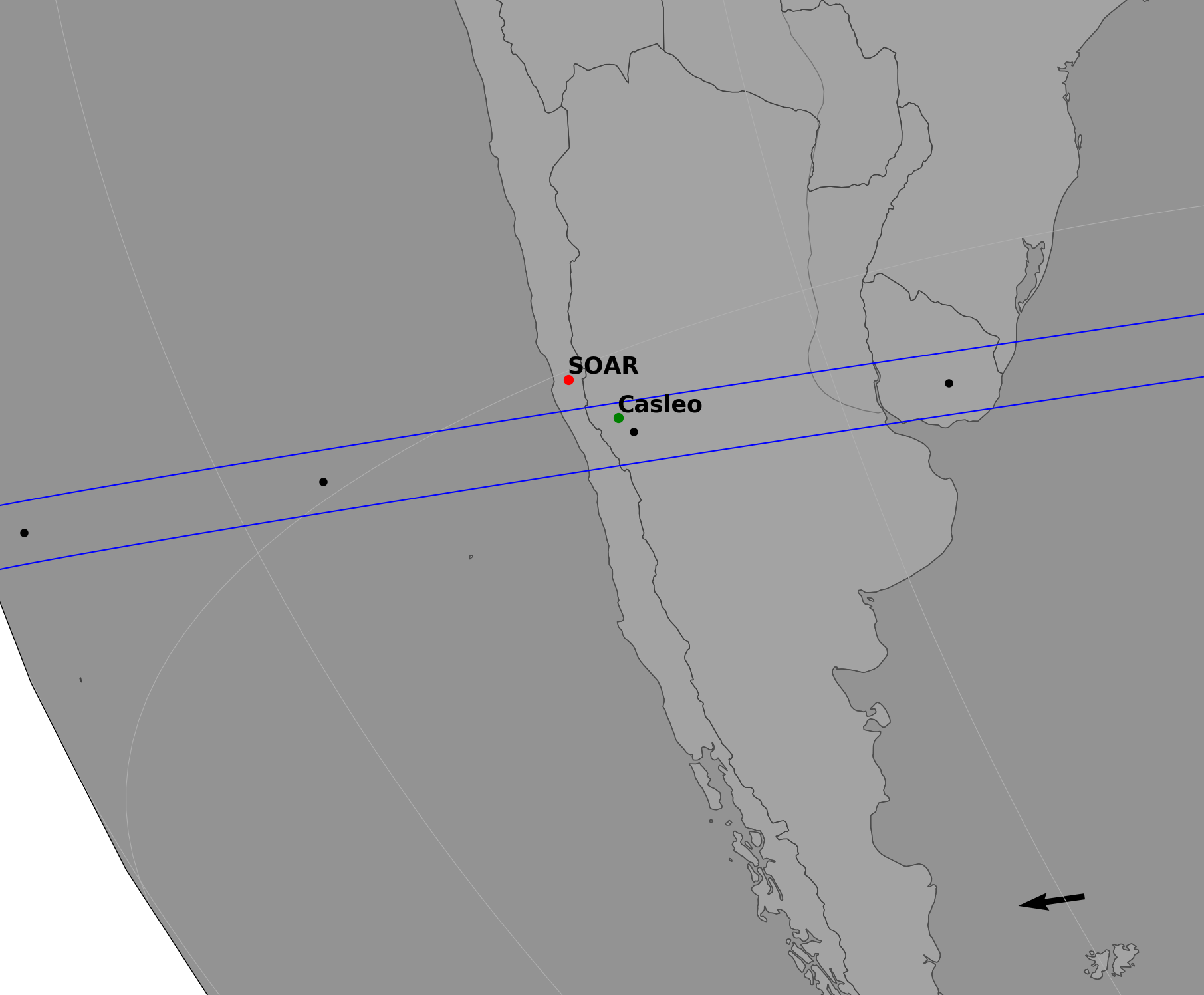}}
\caption{Post-fit maps of Phoebe's occultations. The green dots are the sites that got a positive detection, while the red ones are sites where the detection was negative. The white dot site on the August 13 occultation (e) observed the occultation but with a very low SNR. The blue lines show the path of the shadow with the projected diameter of Phoebe. The black dots show the centre of the shadow separated by one minute. \textcolor{red}{The big black dots show the position of the shadow for the geocentric closest approach between satellite and star}. The arrows on the right low corner show the direction of the shadow's movement.}
\label{Fig:Occ-Phoebe-map}
\end{figure*}

\autoref{Fig:Occ-Phoebe-map} shows the post-reduction occultation maps. \textcolor{red}{The typical difference between the predicted and observed occultations was smaller than Phoebe's radius.} It also shows the location of the sites that attempted to observe the events. In green we have the sites where the occultation was positive, in red the sites with no detection, and in white the site whose observation should be positive, but the low signal-to-noise ratio (SNR) in combination with a long exposure did not allowed the detection of the occultation. \autoref{Tab:Occ-Phoebe-observers} shows the circumstances of the observations, telescopes and detector used for each site on each event.

\begin{table*}
 \centering
  \caption{Circumstances of observation for all observing stations. \label{Tab:Occ-Phoebe-observers}}
  \begin{tabular}{@{}lccccc}
  \hline
          & Longitude & Telescope & Camera & Detection &   \\
     Site & Latitude  & Aperture  & Exp/Cycle time & Ingress & Observers\\          
          & Altitude  & f-ratio & Time-Stamp & Egress    & \\          
\hline
\multicolumn{6}{c}{2017 July 06}\\
\hline
Hamamatsu     & 137\degr44\arcmin23\farcs0 E &    Schmidt Cass     & WAT-120+ & Positive  & Minoru Owada\\
    Japan    & 34\degr43\arcmin07\farcs0 N & 25 cm & 0.534 s  & 16:03:59.55$\pm 0.04$ s &  \\
& 17 m   &  f/9.3   &    GHS-OSD    & 16:04:11.03$\pm 0.04$ s &  \\
\\
    Miharu    & 140\degr26\arcmin04\farcs2 E &    Newtonian     & WAT-120 & Positive  & Katsumasa Hosoi\\
    Japan    & 37\degr25\arcmin36\farcs7 N & 13 cm & 0.534 s  &  16:04:00.83$\pm 0.1$ s &  \\
            & 274 m       &     f/5      &    GHS-OSD    & 16:04:02.07$\pm 0.1$ s &  \\
\hline
\multicolumn{6}{c}{2018 June 19}\\
\hline
 Cerro Pachon  & 70\degr44\arcmin21\farcs7 W & SOAR  & Raptor & Positive  & J. I. B. Camargo\\
 Chile  & 30\degr14\arcmin16\farcs9 S & 420 cm & 0.2 s & 04:37:49.844 $\pm$ 0.009 s & A. R. Gomes Júnior \\
 & 2693.9 m   & f/16  & GPS  & 04:38:01.343 $\pm$ 0.007 s &  \\
\\
 La Silla & 70\degr44\arcmin20\farcs18 W & Danish  & Lucky Imager & Negative  & Justyn Campbell-White \\
 Chile & 29\degr15\arcmin21\farcs27 S & 154 cm & 0.1 s & beg: 04:35:01.05 & Sohrab Rahvar \\
 & 2336 m & f/8.6  & GPS-connected NTP  & end: 04:41:14.07 &  Colin Snodgrass\\
\\
La Silla & 70\degr44\arcmin21\farcs8 W & TRAPPIST-S  & FLI PL3041-BB & Negative & Emmanuel Jehin \\
 Chile & 29\degr15\arcmin16\farcs6 S & 60 cm & 3.0/4.13 s & beg: 04:24:01.82 &  \\
 & 2317.7 m & f/8 & NTP & end: 04:47:19.71 &  \\
\\
 Foz do Iguaçu & 54\degr35\arcmin37\farcs50 W & Celestron  & Raptor & Negative  & Daniel I. Machado\\
  Brazil    & 25\degr26\arcmin05\farcs36 S & 28 cm & 5.0 s & beg: 04:31:32.15 &  \\
& 184.8 m   & f/10 & GPS180PEX, Meinberg  & end: 04:46:27.15 &  \\
\hline
\multicolumn{6}{c}{2018 June 26}\\
\hline
 Penrith Obs. & 150\degr44\arcmin29\farcs9 E & Ritchey-Chretien & Grasshopper Express & Positive  & Tony Barry\\
Australia  & 33\degr45\arcmin43\farcs5 S & 60 cm & 1.0 s & 18:31:20.1 $\pm$ 0.1 s & Ain De Horta \\
& 58 m   & f/10  & ADVS  & 18:31:21.3 $\pm$ 0.1 s & David Giles \\
&    &   &   &  & Rob Horvat \\
\hline
\multicolumn{6}{c}{2018 July 03}\\
\hline
 Rockhampton & 150\degr30\arcmin01\farcs6 E & SCT  & Watec 910BD & Positive  & Stephen Kerr\\
Australia  & 23\degr16\arcmin10\farcs1 S & 30 cm & 1.28 s & 13:37:51.8 $\pm$ 0.2 s &  \\
& 50 m   & f/10  & IOTA-VTI & 13:37:56.9  $\pm$ 0.2 s &  \\
\hline
\multicolumn{6}{c}{2018 August 13}\\
\hline
  Yass & 148\degr58\arcmin35\farcs14 E & Planewave CDK20 & QHY174M-GPS  & Positive  & William Hanna\\
 Australia  & 34\degr51\arcmin51\farcs17 S & 50.8 cm & 2.0 s & 12:52:45.3 $\pm$ 0.7 s &  \\
 & 535 m   & f/4.4  & in-camera GPS & 12:53:08.1 $\pm$ 0.7 s &  \\
\\
Penrith Obs. & 150\degr44\arcmin29\farcs9 E & Ritchey-Chretien & SBIG STT-8300 & Bad SNR  & Tony Barry\\
Australia  & 33\degr45\arcmin43\farcs5 S & 60 cm & 8.0/9.0 s & beg: 12:31:53.0 & Ain De Horta \\
& 58 m   & f/10  & NTP & end: 12:54:32.0 & David Giles \\
&    &   &   &  & Rob Horvat \\
&    &   &   &  & Darren Maybour \\
\hline
\multicolumn{6}{c}{2019 June 07}\\
\hline
 Casleo  & 69\degr17\arcmin44\farcs9 W & Jorge Sahade  & Versarray 2048B, Roper Scientific & Positive  & Luis A. Mammana\\
 Argentina  & 31\degr47\arcmin55\farcs6 S & 215 cm & 0.5/2.0 s & 03:54:22.6 $\pm$ 0.9 s &   Eduardo F. Lajús\\
 & 2552 m   & f/8.5  & GPS  & 03:54:32.5 $\pm$ 0.6 s &  \\
\\
 Cerro Pachon  & 70\degr44\arcmin21\farcs7 W & SOAR  & Raptor & Negative  & J. I. B. Camargo\\
 Chile  & 30\degr14\arcmin16\farcs9 S & 420 cm & 0.4 s & beg: 03:48:44.4 & A. R. Gomes Júnior \\
 & 2693.9 m   & f/16  & SOAR & end: 04:05:02.8 &  \\
\hline
\end{tabular}
Details about the Danish instrument can be found in \cite{Skottfelt2015} and for the TRAPPIST one in \cite{Jehin2011}.
\end{table*}

Many of the observations were made using video cameras. The procedures of the reduction of video observations are similar to those used by \cite{Rossi2016}. Each individual frame from the video is converted to FITS format, with a frame rate used by the observer, using our own code which uses \textsc{astropy}\footnote{http://www.astropy.org} and \textsc{ffmpeg}\footnote{http://ffmpeg.org/} features. All frames were checked to verify the presence of duplicated fields or missing frames in the data set.

To recover the individual exposure of each observation, for example the 0.534 second observed in the 2017 event, a group of frames corresponding to the same exposure was considered, with the first and last frames of each sequence being removed to avoid interlace problems between different exposures. The average of the remaining frames was obtained to represent an individual image. The mid-exposure time of each image was verified by comparing the extracted time to the time printed on the frames of the video, given by a video time inserter. For a better understanding of the procedure, refer to \cite{Rossi2016}.

\section{Light Curve Reduction}
\label{Sec:Curves}

\begin{figure*}
\begin{centering}
\includegraphics[width=\linewidth]{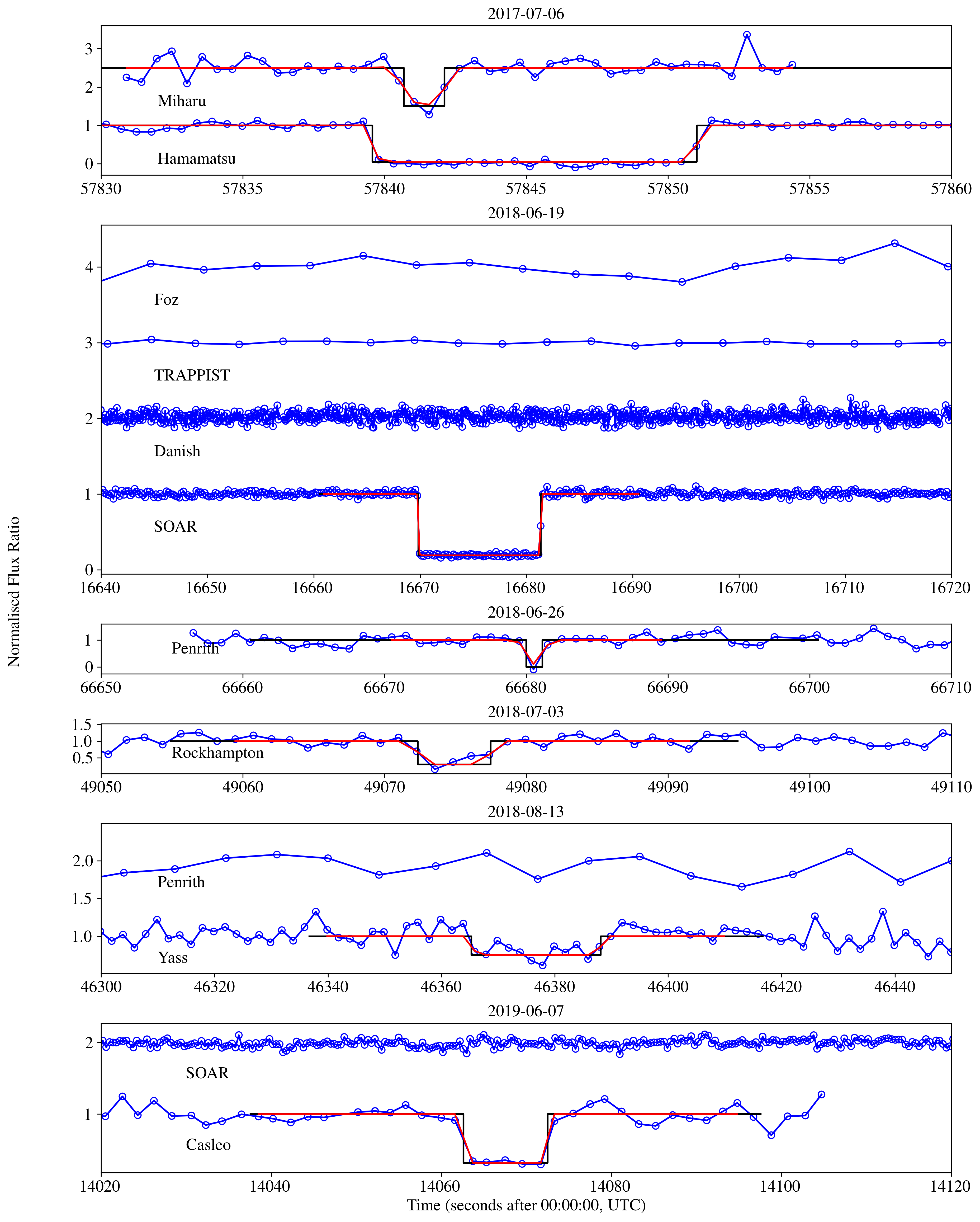}
\caption{Normalised light curves from the six stellar occultations by Phoebe. Some curves are vertically shifted for better visualisation. The blue lines show the normalised flux ratio of Phoebe plus occulted star relative to the calibration star. The solid black lines are the best fit of the square-well model to the data. Red lines are the square-well model convoluted with the Fresnel diffraction, the star diameter, and the applied exposure time.}
\label{Fig:Occ-Phoebe-06jul2017-fluxratio}
\end{centering}
\end{figure*}

For each observation, the flux of the occulted star was determined by a differential aperture photometry using the Package for Reduction of Astronomical Images Automatically \citep[\textsc{praia},][]{Assafin2011}. The light curve obtained was calibrated and normalised by the flux of a calibration star. The final light curves from all positive observations are shown in \autoref{Fig:Occ-Phoebe-06jul2017-fluxratio}.

The procedure of determination of the instants of ingress (disappearance) and egress (reappearance) in the light curves is similar to the one adopted by \cite{BragaRibas2013}, where a square-well model is convoluted with the Fresnel diffraction, the CCD bandwidth, the stellar apparent diameter, and the applied finite exposure time.

From the diameter and parallax of the stars given by Gaia-DR2 \citep{Andrae2018} we determine the diameter projected at Phoebe's distance for the 2017 and 2019 events. For the remaining, the diameter was estimated with the formulae of \cite{vanBelle_1999} by using the magnitudes found in the literature. These values are given in \autoref{Tab:Occ-stars}. The Fresnel scale was between 0.714-0.721 km. For all observations, those effects are negligible compared to the length corresponding to one exposure, thus this last being the main parameter in the determination of the instants of star disappearance (ingress) and reappearance (egress) of the light curves.

The fitting process consists in minimising a classical $\chi^{2}$ function for each light curve, where the free parameter is the ingress or egress instant. \autoref{Fig:Occ-Phoebe-06jul2017-fluxratio} shows all the observed light curves with the corresponding model in red. The ingress and egress instants obtained and respective uncertainties, at 1-$\sigma$ level, are shown in \autoref{Tab:Occ-Phoebe-observers}. For the 2018 August 13 event, the Penrith Observatory is on the shadow path, however, due to the observation conditions, the SNR was too small and the flux drop could not be identified.

\section{The 2017 July 06 occultation}
\label{Sec:Phoebe-2017}

Since Phoebe already has a known shape from \textit{Cassini} observations, we used the 3D model of \cite{Gaskell2013}\footnote{Gaskell (2013): \url{https://space.frieger.com/asteroids/moons/S9-Phoebe}} to fit our chords. As shown by \textit{Cassini}, Phoebe is highly cratered, with the largest one having an estimated size of $\sim100$ km \citep{Porco2005}, with walls that can reach 15 km high, which is significant relative to Phoebe's size. So it is likely that both chords encompass topographical features.

The sub-observer latitude ($\phi$) and longitude ($\lambda$) at the instant of observation (16:04:00 UTC) was determined from \cite{Archinal2018} as $\phi = 22.3\degr$ and $\lambda = 330\degr$. From the pole coordinates of Phoebe \textcolor{red}{($\alpha_p = 356.90\degr$, $\delta_p = 77.80\degr$)}, given by \cite{Archinal2018}, and its ephemeris at the mid-instant of the occultation, from \cite{GomesJunior2016}, we determined the pole position angle as $PA=13.2\degr$.

\begin{figure}
\begin{centering}
\includegraphics[width=\linewidth]{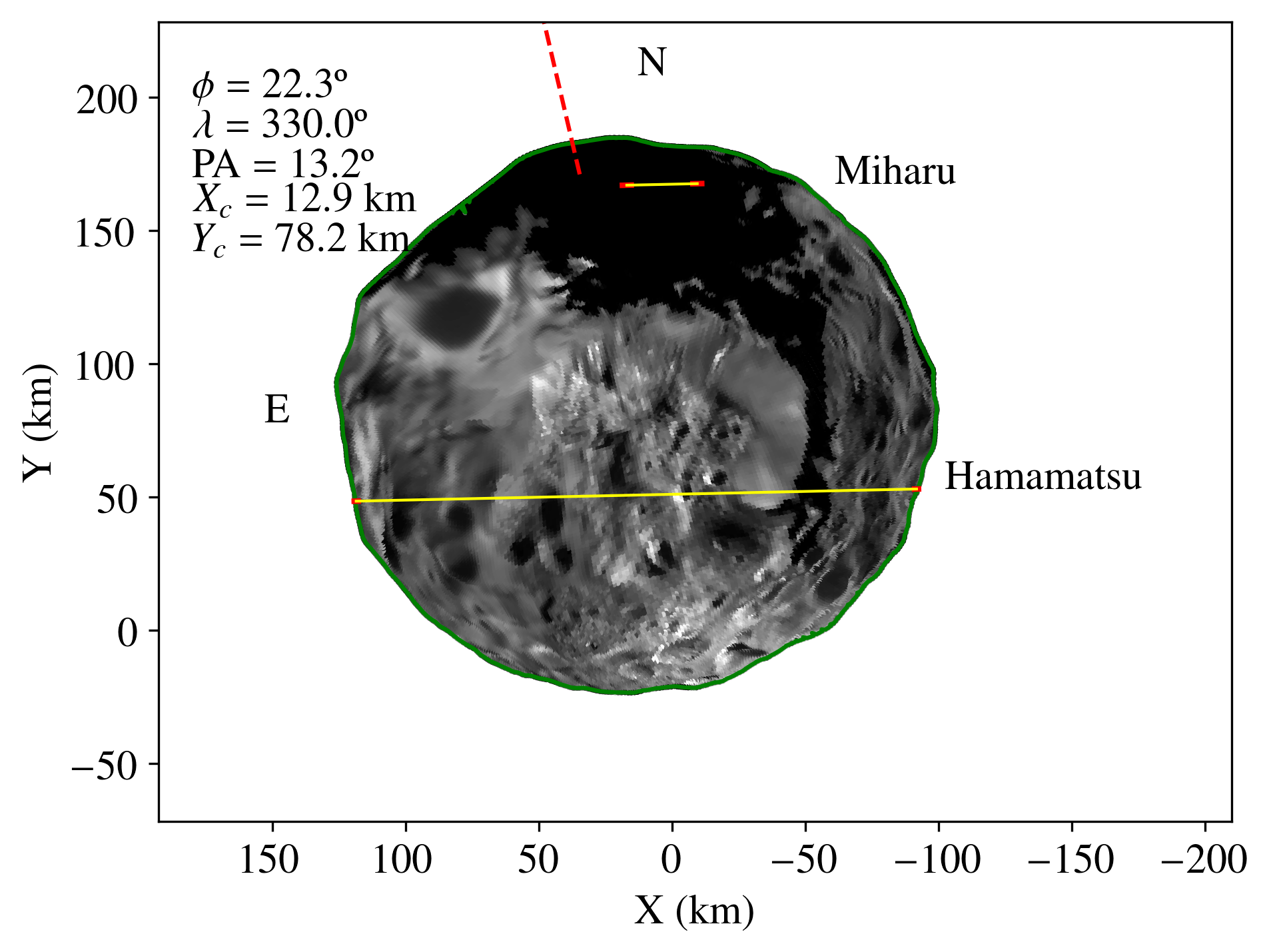}
\caption{Fit of the two chords from the 2017 July 06 occultation to the 3D shape model of Phoebe given by \protect\cite{Gaskell2013}. The sub-observer latitude $\phi$ and longitude $\lambda$ was calculated from the rotation model of \protect\cite{Archinal2018}. $PA$ is the position angle of the north pole of Phoebe. $X_c$ and $Y_c$ are the coordinates of the centre of Phoebe relative to the prediction, taking the star coordinates from Gaia DR2. The yellow lines show the observed chords and the red segments show their error bars. The dashed red line shows the direction of the North Pole of Phoebe.}
\label{Fig:Occ-Phoebe-3d-330}
\end{centering}
\end{figure}

\autoref{Fig:Occ-Phoebe-3d-330} shows the fit from the chords to the 3D shape model of Phoebe using the presented orientation. It is possible to see that the Miharu chord is close to the North Pole of the object. In green, the limb of Phoebe projected on the sky plane is highlighted. The centre was determined by fitting the Hamamatsu chord. Since it is in a central region, this chord is expected to be located in a region that was better observed by \textit{Cassini} than Miharu's. The surface brightness texture presented in the image comes from \textit{Cassini} observations given by \cite{Gaskell2013}. The regions in black were not observed by \textit{Cassini} or were always in darkness.

It is possible to see in \autoref{Fig:Occ-Phoebe-3d-330} that the Miharu chord is located very far from the projected limb of the 3D shape model, while the Hamamatsu chord fits very well. Regions observed by \textit{Cassini} in the continuation of the chord were not detected in the occultation.

In preparation for the \textit{Cassini} mission, \cite{Bauer2004} determined the rotational period of Phoebe \textcolor{red}{as $9.2735 \pm 0.0006$ h}. Propagating it from 2004 to the occultation time, it represents an uncertainty in the sub-observer longitude of $\pm 305\degr$. That means that the nominal sub-observer longitude calculated from \cite{Archinal2018} is probably wrong. Since the sub-observer latitude and position angle are better determined, the goal now is to determine the sub-observer longitude that better fits the 3D shape model and occultation chords.

To retrieve the correct sub-observer longitude, a $\chi^2$ analysis was done for all longitudes, ranging from $0\degr$ to $360\degr$ with a $0.5\degr$ step. Since the Hamamatsu chord probed the central region, where the 3D model has a better spatial resolution, only this chord was used to fit the object's centre ($X_c$, $Y_c$). With this condition satisfied, we calculated the total $\chi^2$ with the chord extremities using \autoref{Eq:Occ-Phoebe-chi-quadrado}:
\begin{equation}
\chi^2 = \sum \dfrac{(R_c - R_l)^2}{\sigma^2}
\label{Eq:Occ-Phoebe-chi-quadrado}
\end{equation}
where $R_c$ is the radial distance of each chord extremity from the geometric centre given by ($X_c$, $Y_c$) and $R_l$ is the radial distance of the limb of the 3D shape model at the same direction as $R_c$. The uncertainty $\sigma$ of each point was 0.7 km for Hamamatsu and 1.6 km for Miharu.

\begin{figure}
\begin{centering}
\includegraphics[width=\linewidth]{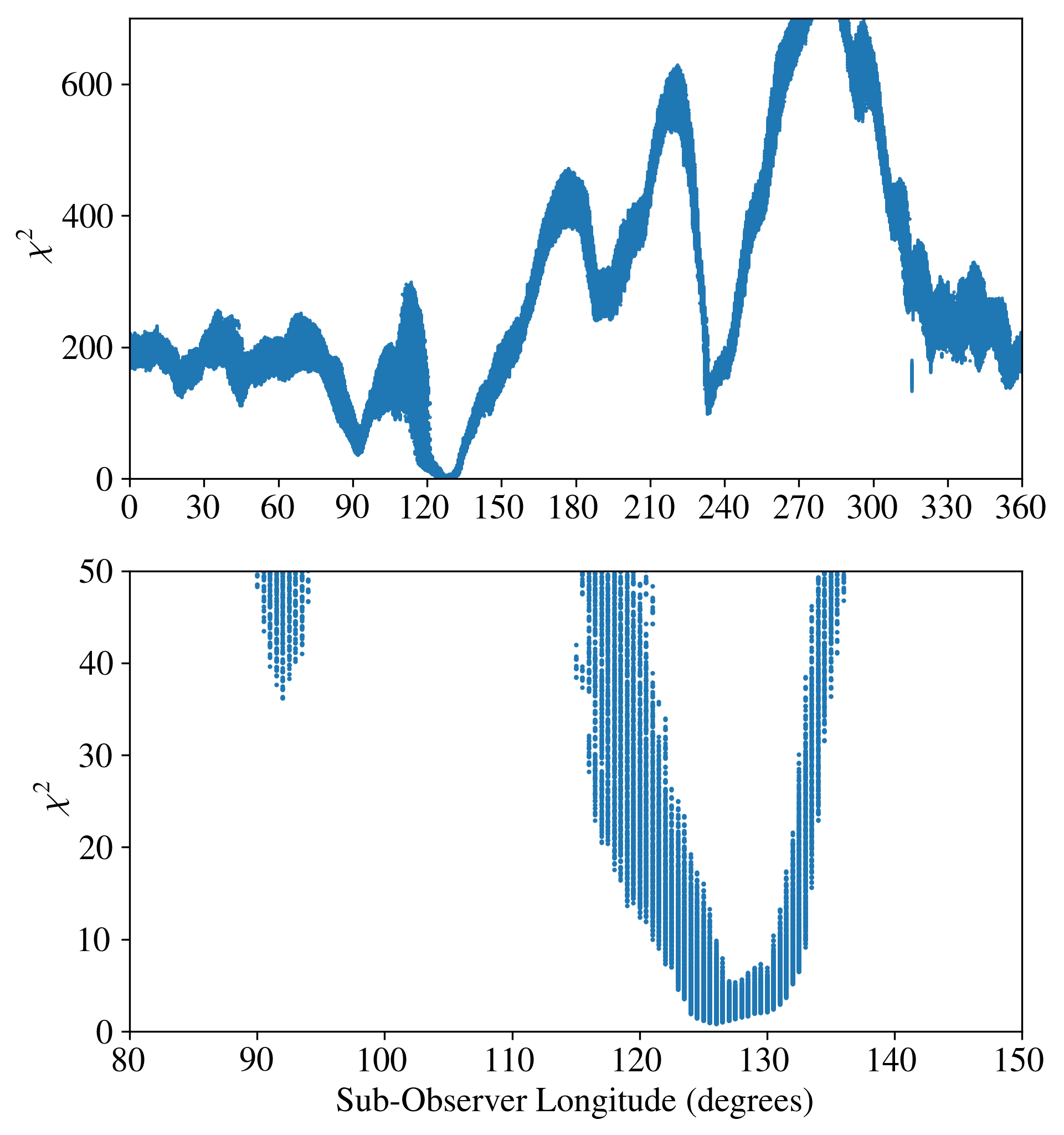}
\caption{Distribution of $\chi^2$ over longitude (see discussion in \autoref{Sec:Phoebe-2017}). Upper panel: chi-square over all longitudes. Bottom panel: zoom in the region of local minima.}
\label{Fig:Occ-Phoebe-chi-quadrado}
\end{centering}
\end{figure}

\autoref{Fig:Occ-Phoebe-chi-quadrado} shows the distribution of $\chi^2$ by longitude. It is possible to see that the $\chi^2$ varies from almost zero up to 700. For comparison, in \autoref{Fig:Occ-Phoebe-3d-330} $\chi^2$ is equal to 201.

Although the Miharu chord crosses a region that was not well observed by \textit{Cassini}, it is expected the \cite{Gaskell2013} shape model to be realistic enough considering the relatively higher error of this chord. Thus, it is difficult to establish which value of $\chi^2$ is acceptable. Because of this, we analyse the local minima presented in \autoref{Fig:Occ-Phoebe-chi-quadrado}.

\begin{figure}
\begin{centering}
\includegraphics[width=\linewidth]{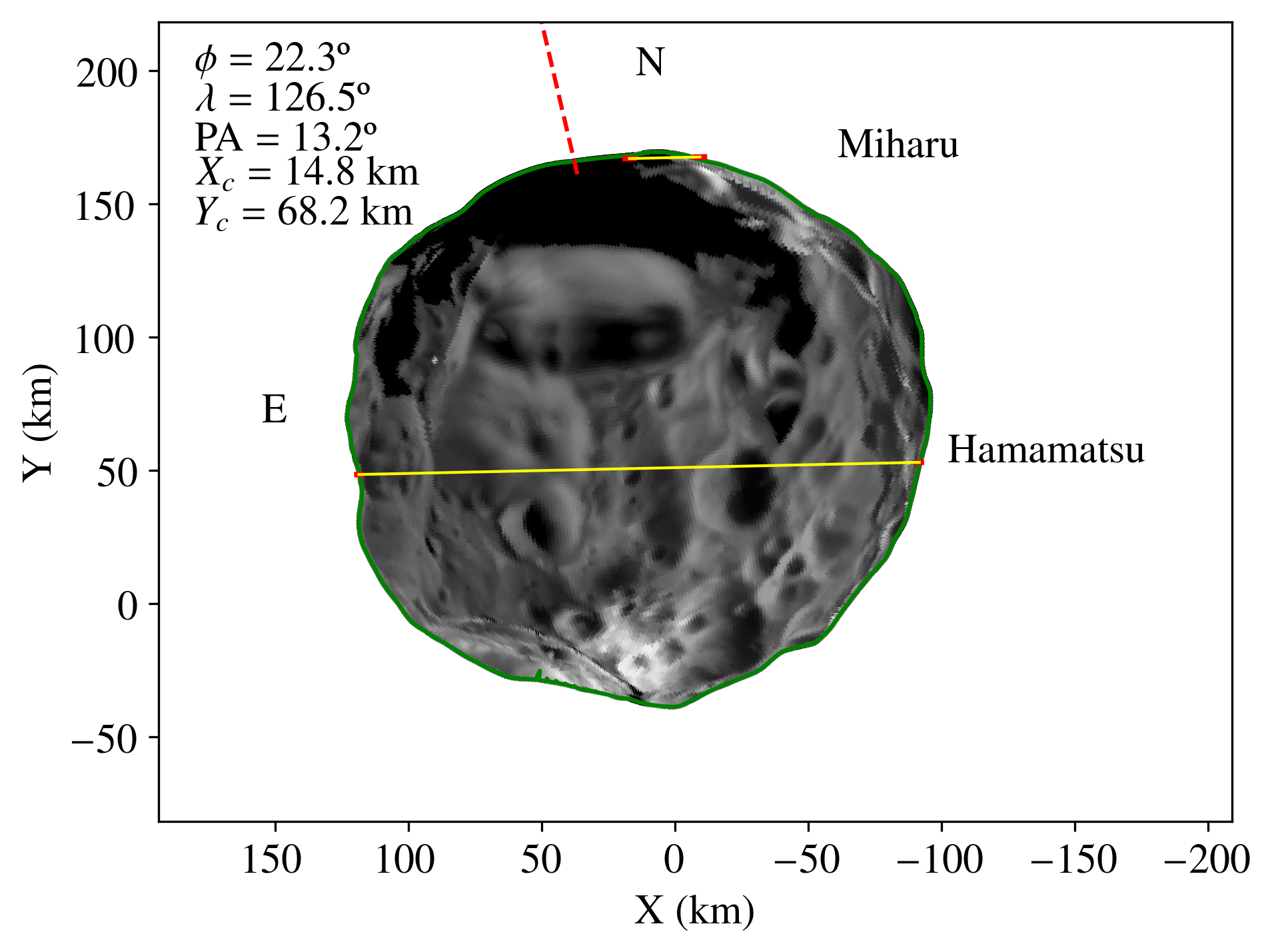}
\caption{Similar to the \autoref{Fig:Occ-Phoebe-3d-330}, but fitting the chords to the 3D shape model using the sub-observer longitude of $126.5\degr$.}
\label{Fig:Occ-Phoebe-3d-127}
\end{centering}
\end{figure}

\begin{figure}
\begin{centering}
\includegraphics[width=\linewidth]{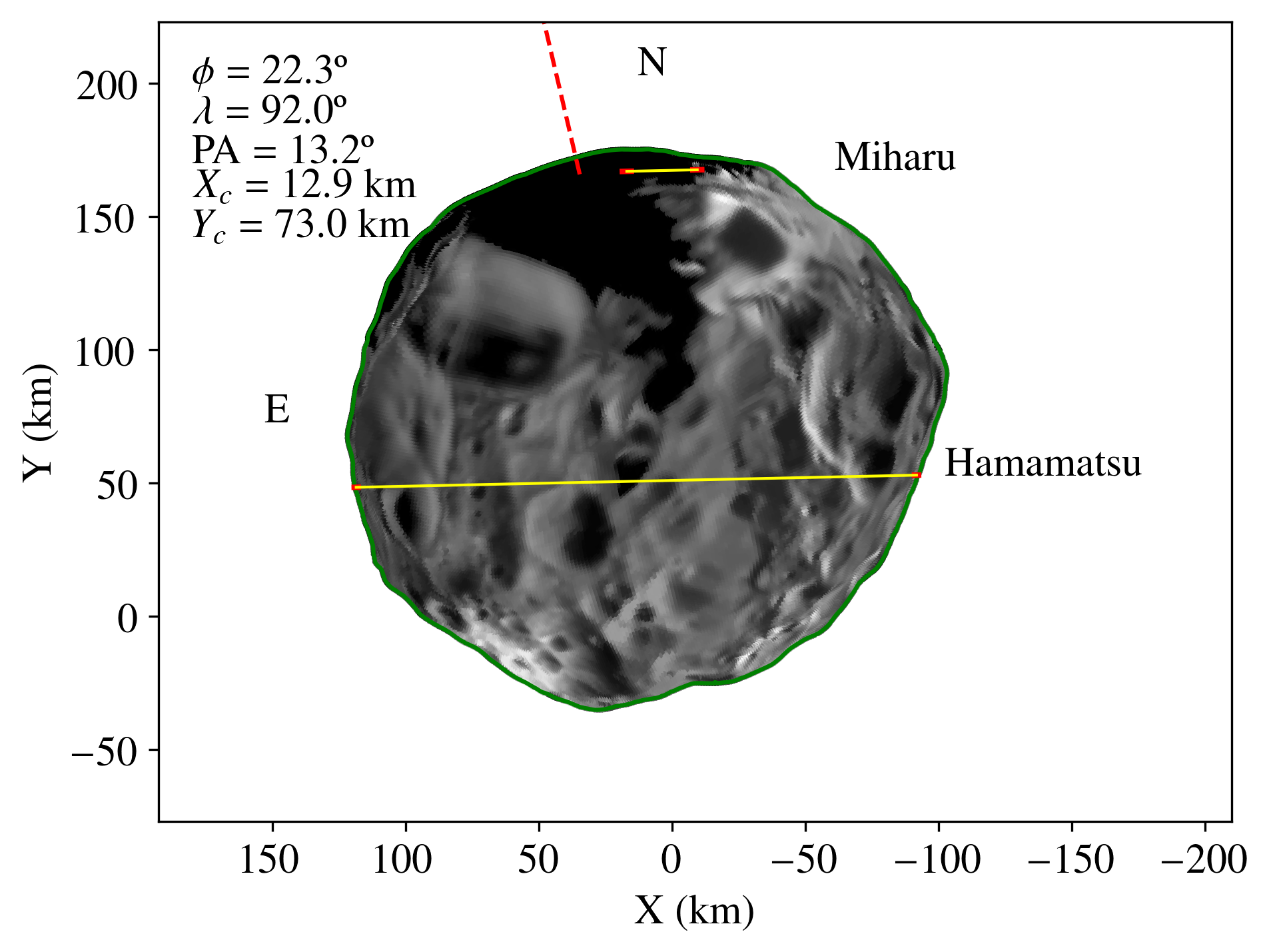}
\caption{Similar to the \autoref{Fig:Occ-Phoebe-3d-330}, but fitting the chords to the 3D shape model using the sub-observer longitude of $92\degr$.}
\label{Fig:Occ-Phoebe-3d-92}
\end{centering}
\end{figure}

The most prominent minimum is at the longitude $126.5\degr$ with a $\chi^2 = 0.85$. \autoref{Fig:Occ-Phoebe-3d-127} shows the best fit in this case. It is possible to see that the western contact point of Miharu chord seems to be in a region that was observed by \textit{Cassini}, though it was a region observed in lower resolution. The second minimum in the \autoref{Fig:Occ-Phoebe-chi-quadrado} is located at the longitude $92.0\degr$ with a $\chi^2 = 36.2$. \autoref{Fig:Occ-Phoebe-3d-92} shows the best fit in this case.

In \autoref{Tab:Occ-Phoebe-127} we present the limits of values ($\lambda$, $X_c$, $Y_c$) within 1-$\sigma$ ($\chi^2_{min}+1$) for the two main chi-square minima of \autoref{Fig:Occ-Phoebe-chi-quadrado}. The radial residuals of the Miharu chord are also presented, separated by its western and eastern contact points. Thus, we can infer possible topographic characteristics, if any.

\begin{table}
\caption{\label{Tab:Occ-Phoebe-127} Results of $\lambda$, $X_c$, $Y_c$ for the two chi-square minima of the \autoref{Fig:Occ-Phoebe-chi-quadrado}. Miharu W refers to radial residuals between the chord and the 3D model to the western contact point of the Miharu chord and Miharu E to the eastern contact point.}
\begin{centering}
\begin{tabular}{ccccc}
\hline  \hline
$\lambda$  & $X_c$ & $Y_c$ & Miharu W & Miharu E \tabularnewline
(\degr) &  (km) &  (km) &  (km) &  (km) \tabularnewline
\hline
$126.5\pm3.5$ & $14.8\pm0.7$ & $68.2\pm1.1$ & $0.8\pm1.2$ & $-1.2\pm1.0$ \tabularnewline 
$92.0\pm1.0$ & $12.9\pm0.2$ & $73.0\pm0.1$ & $-4.9\pm0.1$ & $-8.2\pm0.1$ \tabularnewline 
\hline
\end{tabular}
\par \end{centering}
\end{table}

The results of \autoref{Tab:Occ-Phoebe-127} show that for the $\lambda=126.5\degr$ solution the western contact point of the Miharu chord agrees with the projected limb of the 3D model, while the eastern contact point suggest a slight variation in the direction of the centre of the figure, that could be caused by a crater not taken into account in the shape model.

In the case of $\lambda=92\degr$, the results suggest that both contact points of the Miharu chord would be in craters with 5 to 8 km of difference from the shape model of \cite{Gaskell2013}. \autoref{Fig:Occ-Phoebe-3d-92} shows that some regions observed by \textit{Cassini} west of the chord should have been detected.

For other longitudes, the $\chi^2$ calculated are even less significant, meaning larger differences between the chords and the 3D model of \cite{Gaskell2013}. Because of this, it is not expected to find better results with other longitudes. Considering it all, the sub-observer longitude of $126.5\degr \pm 3.5\degr$ is chosen as our unique solution at the epoch of the 2017 July 06 occultation.

\section{Improved rotational period}
\label{Sec:Phoebe-rotation}

The rotation model of Phoebe from \cite{Archinal2018} predicts the sub-observer longitude at the time of the 2017 July 06 occultation as $\lambda = 330.5\degr$. From the analysis in \autoref{Sec:Phoebe-2017}, we identify a phase difference of $-204.0\pm3.5\degr$ or $+156.0\pm3.5\degr$, representing an error of about half a rotation. Both possibilities are inside the $\pm305\degr$ sub-observer longitude error bar estimate from \cite{Archinal2018} at the occultation epoch.

From \cite{Archinal2018} we calculate the location of the prime meridian (the half great-circle connecting the body's north and south pole defined as $\lambda=0\degr$ in the body-centred reference frame), using $W = 178.58 + 931.639d$ ($W$ is the ephemeris position of the prime meridian and $d$ is the interval in days from J2000), at the epoch of the \textit{Cassini} maximum approach ($W_{\text{cas}}$), where it has the smallest error (Thomas, 2018, priv. comm.), and the instant of occultation ($W_{\text{occ}}$). Then we apply the phase difference determined for $W_{\text{occ}}$ and, keeping $W_{\text{cas}}$ fixed, we fit new parameters.

For the phase difference of $-204.0\degr$ we obtain $W= 247.92 + 931.5963d$, which means a rotational period of \textcolor{red}{$9.27440 \pm 0.00002$ h}. For the $+156.0\degr$ phase difference, we determine $W = 125.56 + 931.6717d$ meaning a rotational period of \textcolor{red}{$9.27365 \pm 0.00002$ h}. We denominate these solutions as "W1" and "W2" respectively.

\section{The 2018 and 2019 stellar occultations}
\label{Sec:Phoebe-2018}

In 2018 and 2019 five other stellar occultations by Phoebe were observed. Only one positive detection was obtained for each event. Because of this, it is not possible to confidently apply the procedure described in \autoref{Sec:Phoebe-2017}. Even so, we still can analyse the possible solutions obtained in \autoref{Sec:Phoebe-rotation} and constrain Phoebe's positions.

We first determined the new longitudes of Phoebe in the direction of the observers for each occultation using the solutions obtained in \autoref{Sec:Phoebe-rotation}. These values are shown in \autoref{Tab:Occ-2018}. Then we fit the chords to the shape model using only the latitude, pole position angle and the 1-sigma interval of longitude to obtain all the possible $X_c$ and $Y_c$ relative to \cite{GomesJunior2016} ephemeris. These values are also shown in \autoref{Tab:Occ-2018}.

\begin{table*}
\caption{Results for the centre of Phoebe from the stellar occultations of 2018 and 2019. For each event the sub-observer longitude ($\phi$), the Position Angle (PA), the sub-observer longitude ($\lambda$) and the ephemeris offset ($X_c$, $Y_c$) are presented. The results are shown for both solutions of the rotation period described in \autoref{Sec:Phoebe-rotation}. 
\label{Tab:Occ-2018}}
\begin{centering}
\begin{tabular}{lcccccccc}
\hline  \hline
Date  & & &  \multicolumn{3}{c}{W1 solution} & \multicolumn{3}{c}{W2 solution} \tabularnewline
 & $\phi$  ($\degr$) & PA  ($\degr$) & $\lambda$ ($\degr$) &  $X_c$ (km) &  $Y_c$ (km) & $\lambda$ ($\degr$) &  $X_c$ (km) &  $Y_c$ (km) \tabularnewline
\hline
 2018 June 19* & 19.8 & 12.8 & $225.5^{+2.5}_{-3.7}$ & -45.5 $\pm$ 1.5 & -25.0 $\pm$ 8.0 & 251.5 $\pm$ 3.7 & -46.0 $\pm$ 1.0 & -40.0 $\pm$ 3.0 \tabularnewline
 2018 June 26 & 20.0 & 12.8 & 86.5 $\pm$ 3.7 & -75.0 $\pm$ 6.5 & -44.5 $\pm$ 1.5 & 112.5 $\pm$ 3.7 & -59.0 $\pm$ 5.5 & -44.0 $\pm$ 1.0 \tabularnewline
 2018 July 03 & 20.1 & 12.8 & 298.5 $\pm$ 3.7 & -40.5 $\pm$ 7.0 & -32.0 $\pm$ 3.0 & 325.0 $\pm$ 3.7 & -42.5 $\pm$ 7.0 & -33.5 $\pm$ 3.5 \tabularnewline
 2018 Aug 13 & 20.8 & 13.0 & 305.5 $\pm$ 3.7 & -42.5 $\pm$ 9.5 & -38 $\pm$ 42 & 335.5 $\pm$ 3.7 & -42.0 $\pm$ 7.5 & -41.5 $\pm$ 40.0 \tabularnewline
 2019 June 07 & 16.5 & 11.7 & 356.0 $\pm$ 4.0 & -108 $\pm$ 12 & -98 $\pm$ 15 & 48.5 $\pm$ 4.0 & -110 $\pm$ 14 & -97 $\pm$ 16 \tabularnewline
\hline
\end{tabular}
\par
\end{centering}
* The expected value for W1 is $225.5\pm3.7$, however longitudes between $228\degr$ and $235\degr$ are improbable, as explained in the text.
\end{table*}

For the June 26 occultation, the centre of Phoebe was well constrained by the negative chords of La Silla and the high SNR chord of SOAR. Given the large size of the SOAR chord, we found that it could not fit the 3D shape model between the longitudes $228\degr$ and $235\degr$ in the 1-sigma level. In this region, the chord is larger than the shape model, given the direction of the chord. We notice that this interval is close to the W1 longitude.

For the events of 2018 June 26 and July 03, there were two possible solutions for the centre, one with the chords located on the north side of the body and another on the south. The north solutions are ($-39\pm24$, $-248\pm4$) and ($-26\pm7$, $-215\pm4$) km for W1 and ($-49\pm22$, $-248\pm2$) and ($-24\pm7$, $-219\pm4$) km for W2, for June 26 and July 03 respectively. However, these occultations happened less than two weeks from the June 19 event, which presents a well constrained position. Large variations in the ephemeris offsets are not expected in such a short time. Because of this, the north solutions were discarded.

The results in \autoref{Tab:Occ-2018} show very precise positions of the centre of Phoebe, except for the 2018 August 13 event. This is due to the chord being close to the centre of Phoebe and its error in the determination of the instants of ingress and egress being larger. The north and south solutions cannot be easily distinguished. In W2, the north and south solutions can be separated, but they are very close to each other. Consequently, we only consider a single solution for W2.

\begin{figure*}
\centering
\subfigure[][W1 solution]{\includegraphics[width=0.48\textwidth]{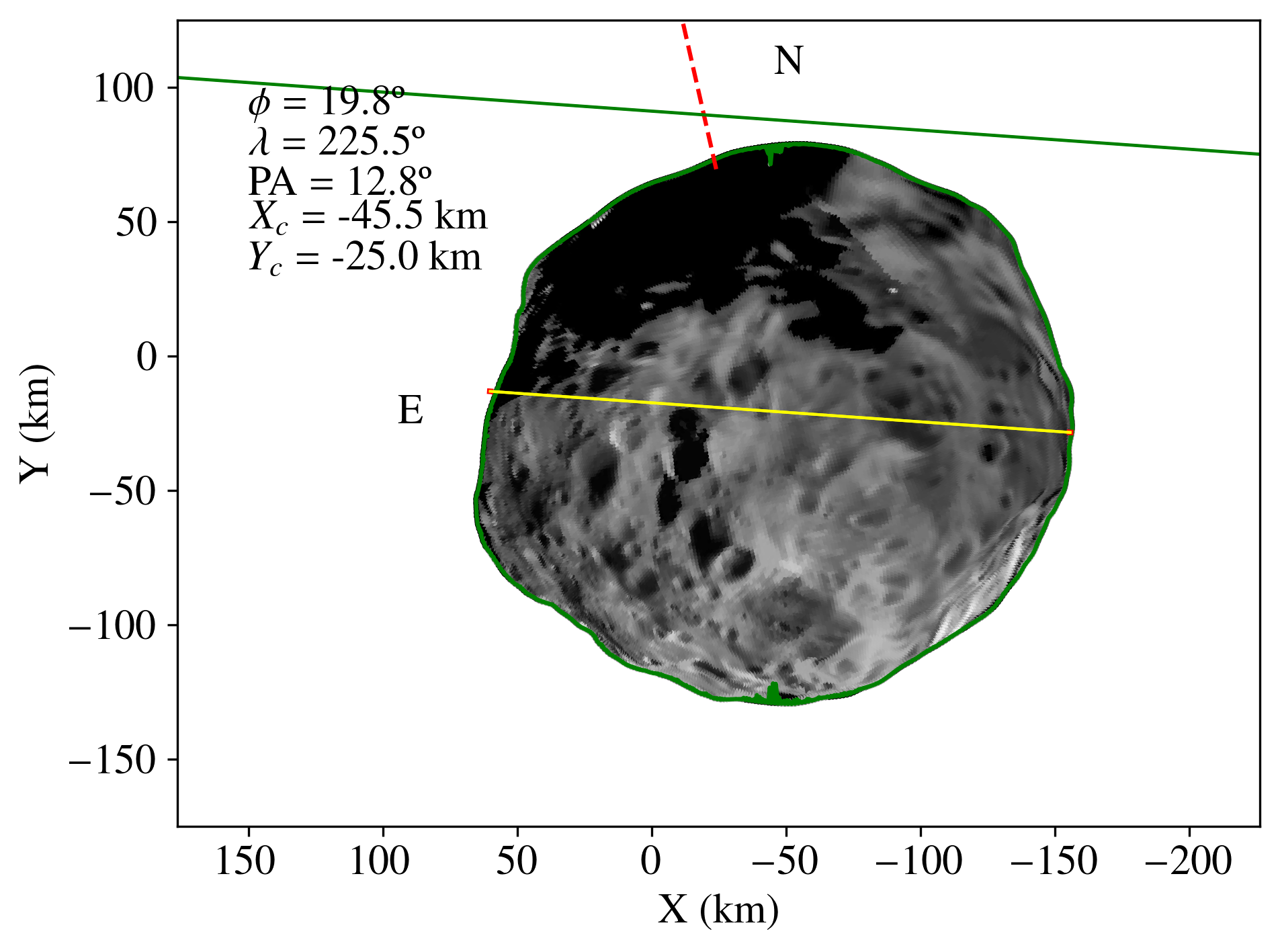}}
\subfigure[][W2 solution]{\includegraphics[width=0.48\textwidth]{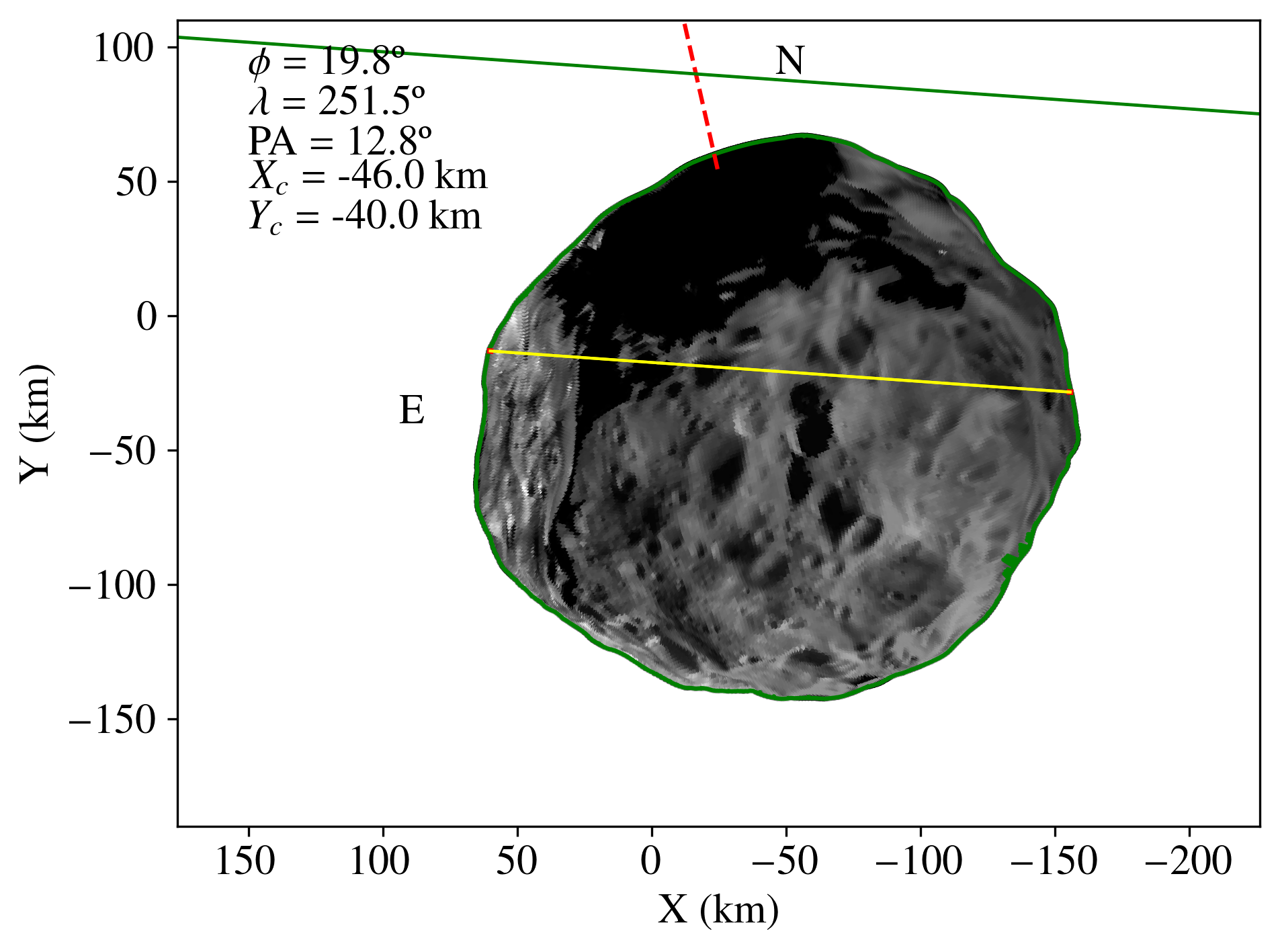}}
\caption{Similar to \autoref{Fig:Occ-Phoebe-3d-330} for the 2018 June 19 occultation. The shapes were obtained using the (a) W1 and (b) W2 rotational period solutions.}
\label{Fig:Occ-Phoebe-jun19}
\end{figure*}

\begin{figure*}
\centering
\subfigure[][W1 solution]{\includegraphics[width=0.48\textwidth]{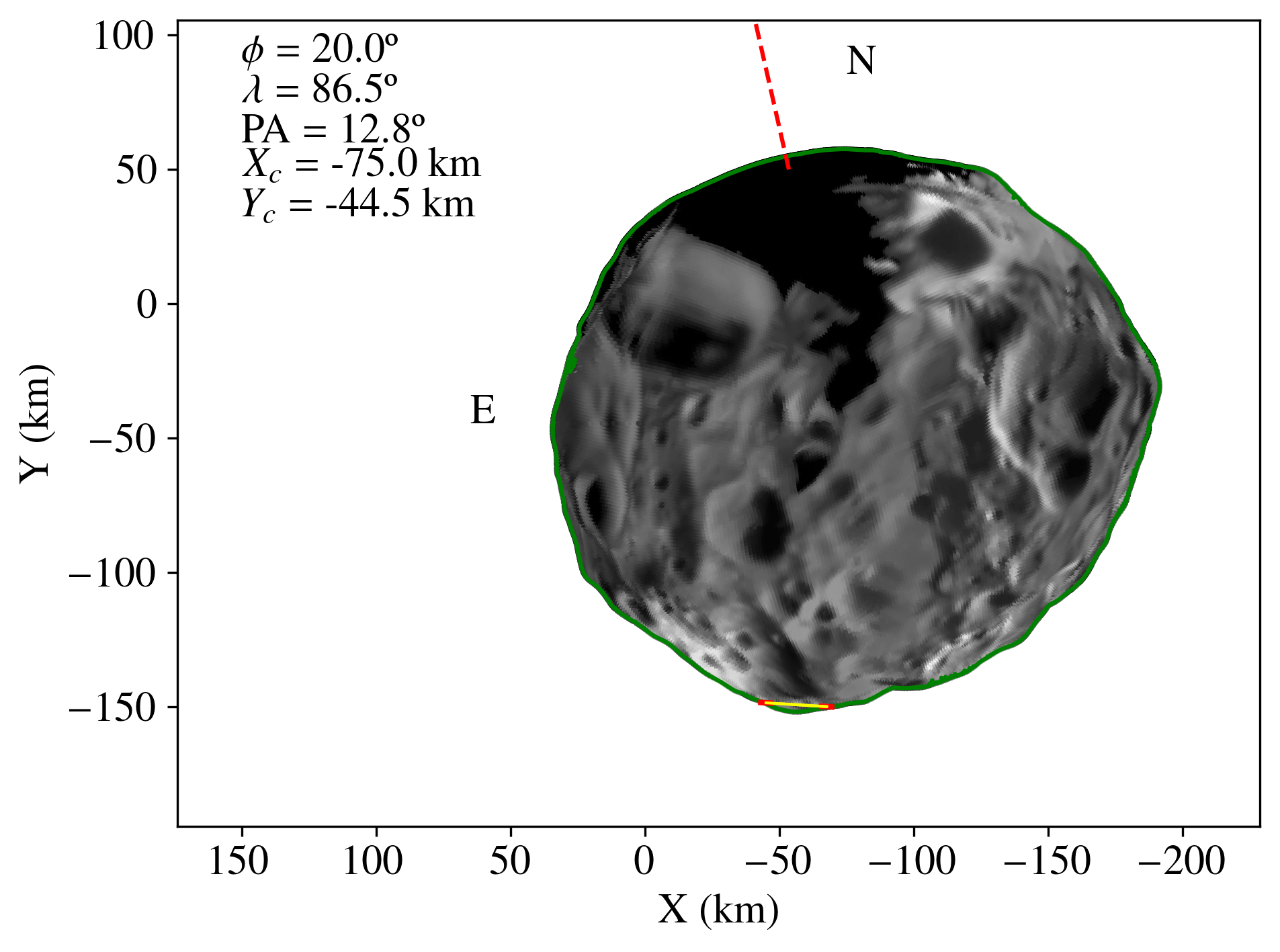}}
\subfigure[][W2 solution]{\includegraphics[width=0.48\textwidth]{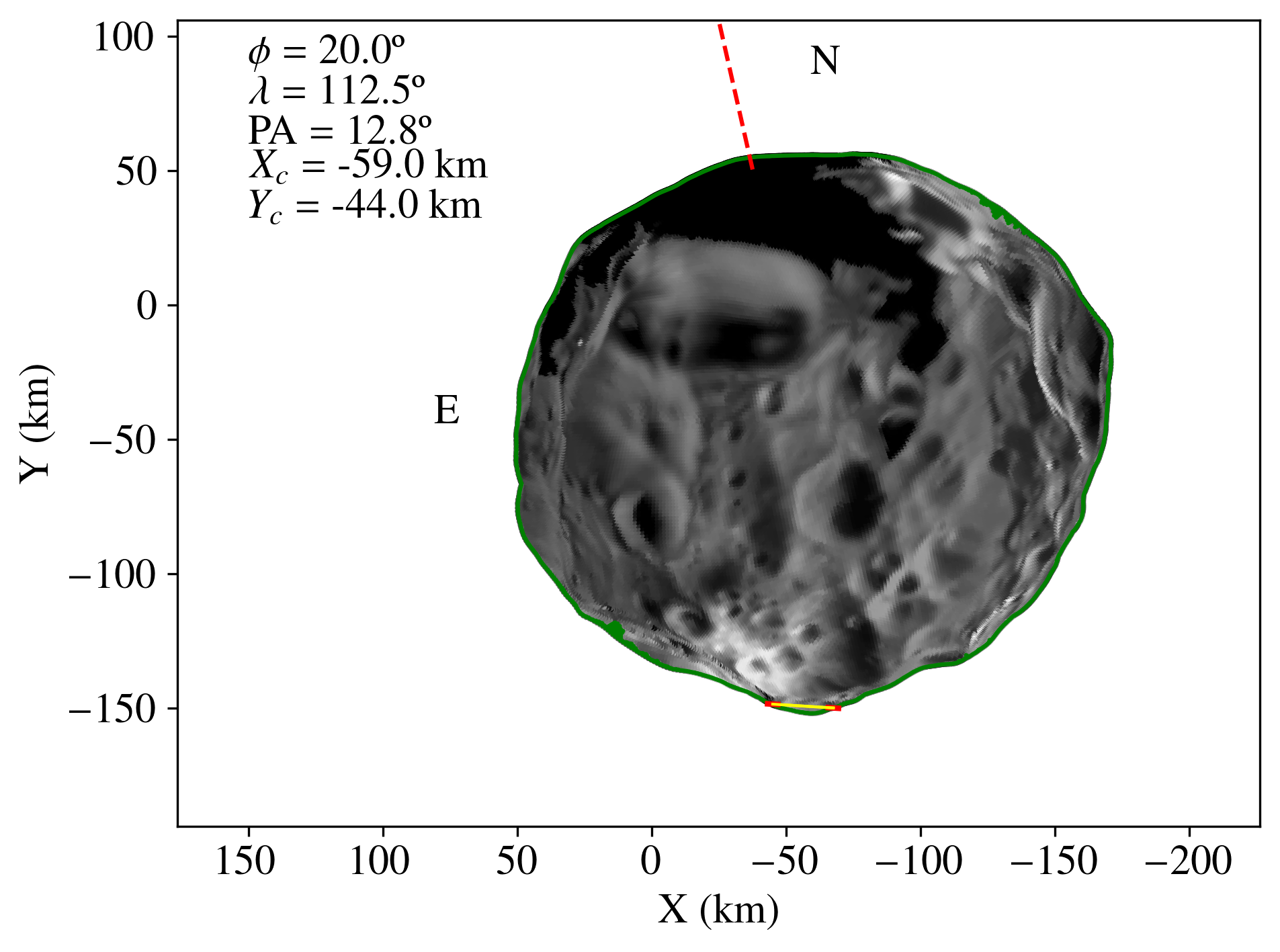}}
\caption{Similar to \autoref{Fig:Occ-Phoebe-jun19} for the 2018 June 26 occultation.}
\label{Fig:Occ-Phoebe-jun26}
\end{figure*}

\begin{figure*}
\centering
\subfigure[][W1 solution]{\includegraphics[width=0.48\textwidth]{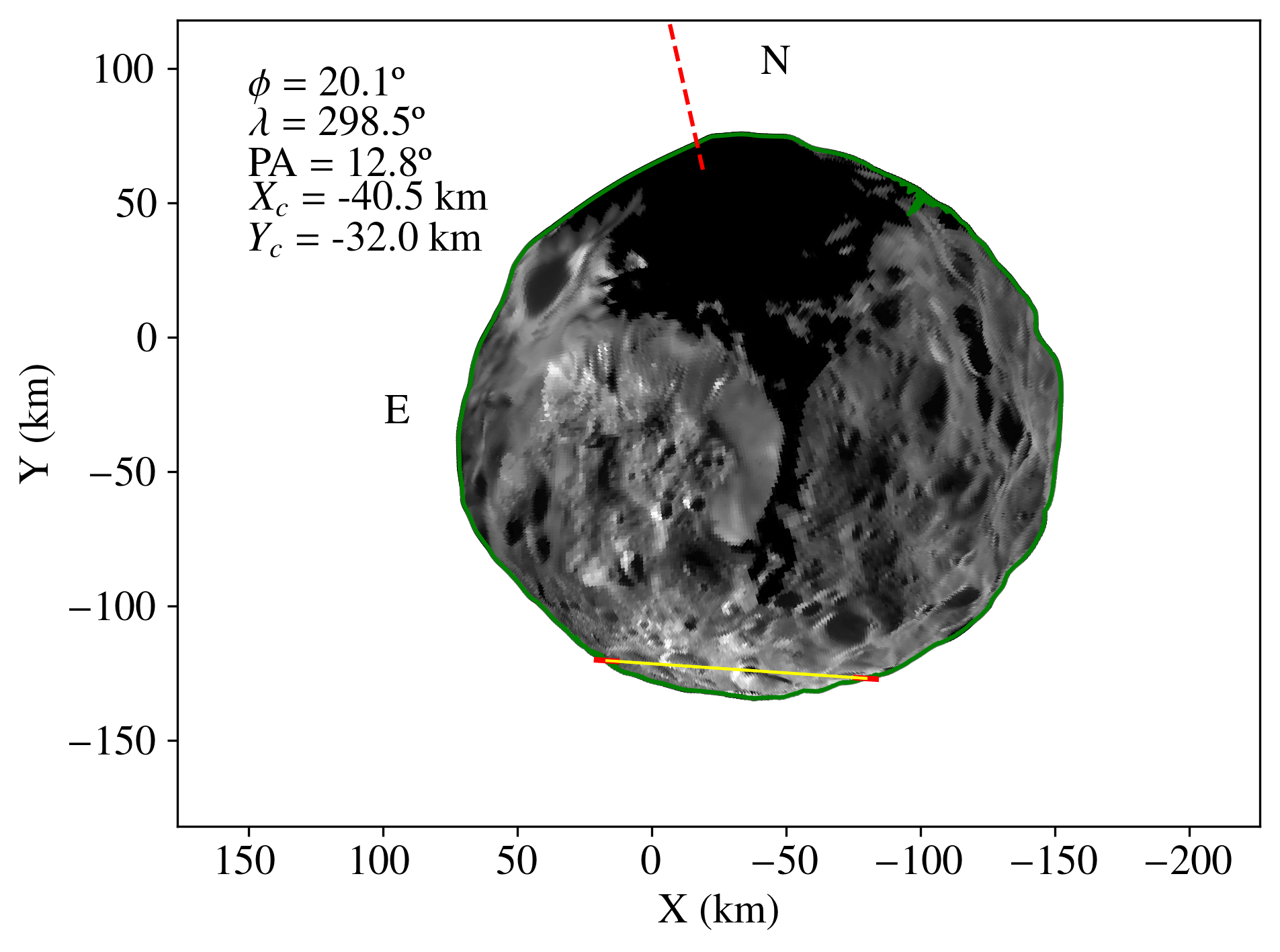}}
\subfigure[][W2 solution]{\includegraphics[width=0.48\textwidth]{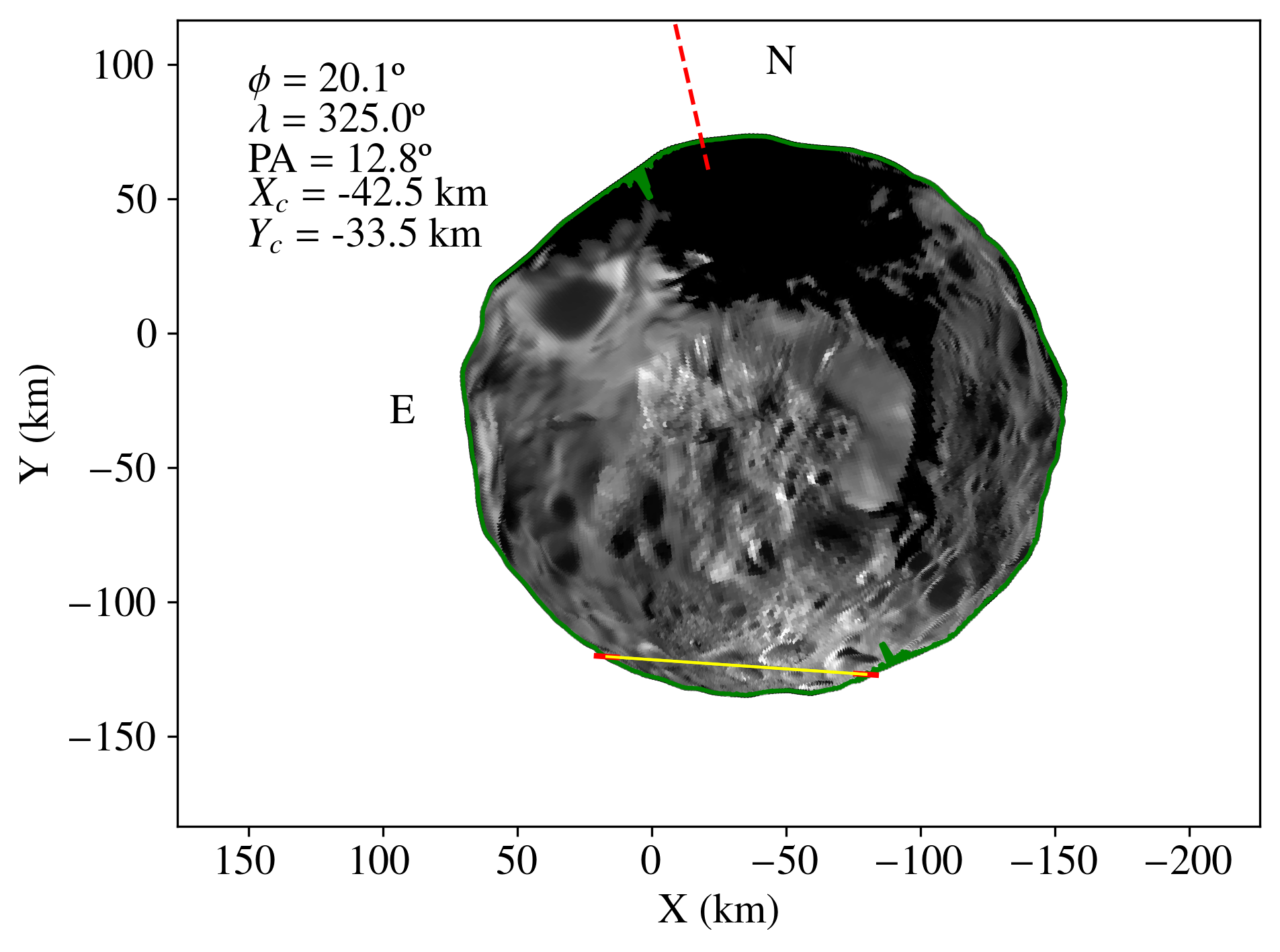}}
\caption{Similar to \autoref{Fig:Occ-Phoebe-jun19} for the 2018 July 03 occultation.}
\label{Fig:Occ-Phoebe-jul03}
\end{figure*}

\begin{figure*}
\centering
\subfigure[][W1 solution]{\includegraphics[width=0.48\textwidth]{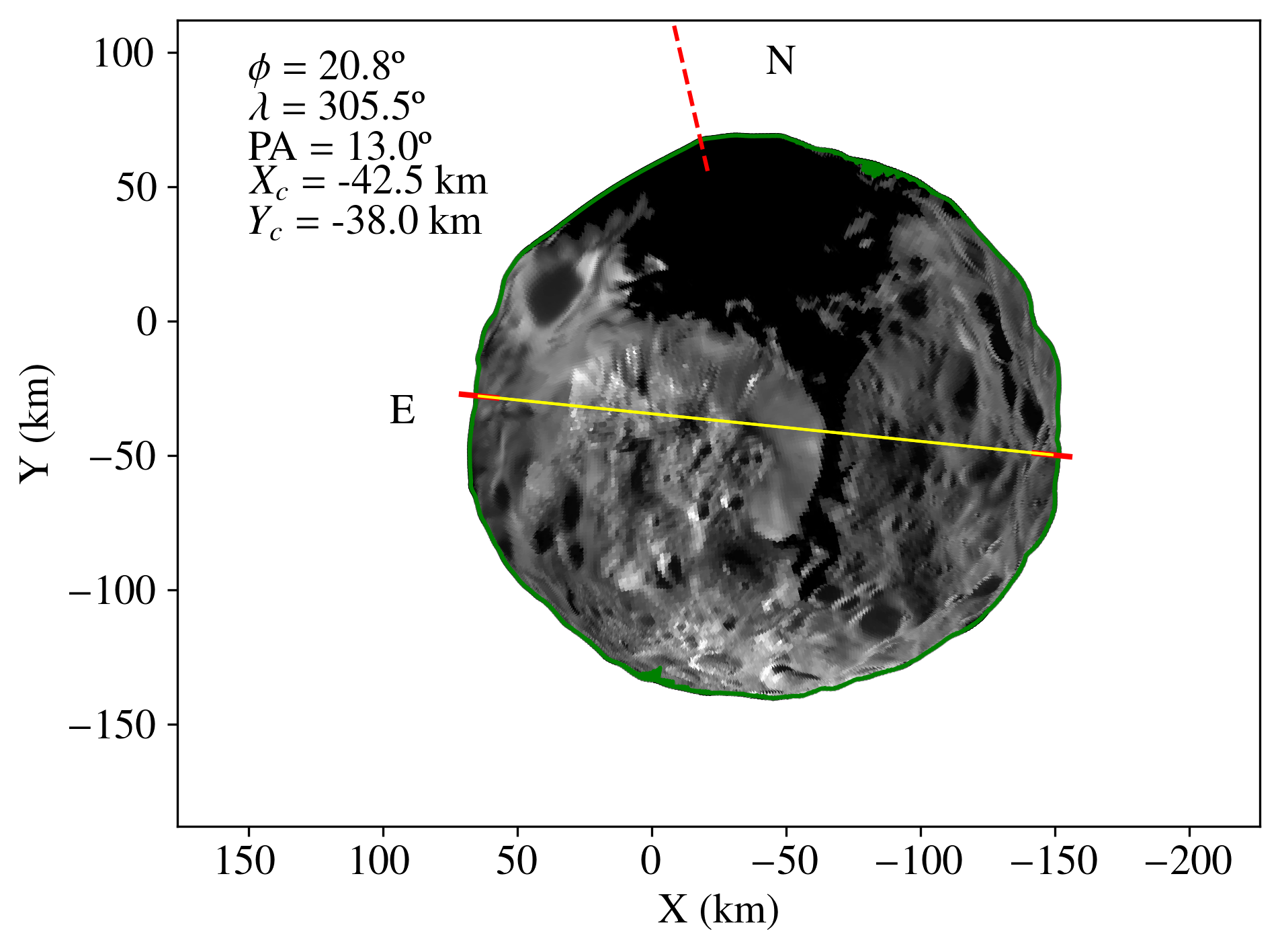}}
\subfigure[][W2 solution]{\includegraphics[width=0.48\textwidth]{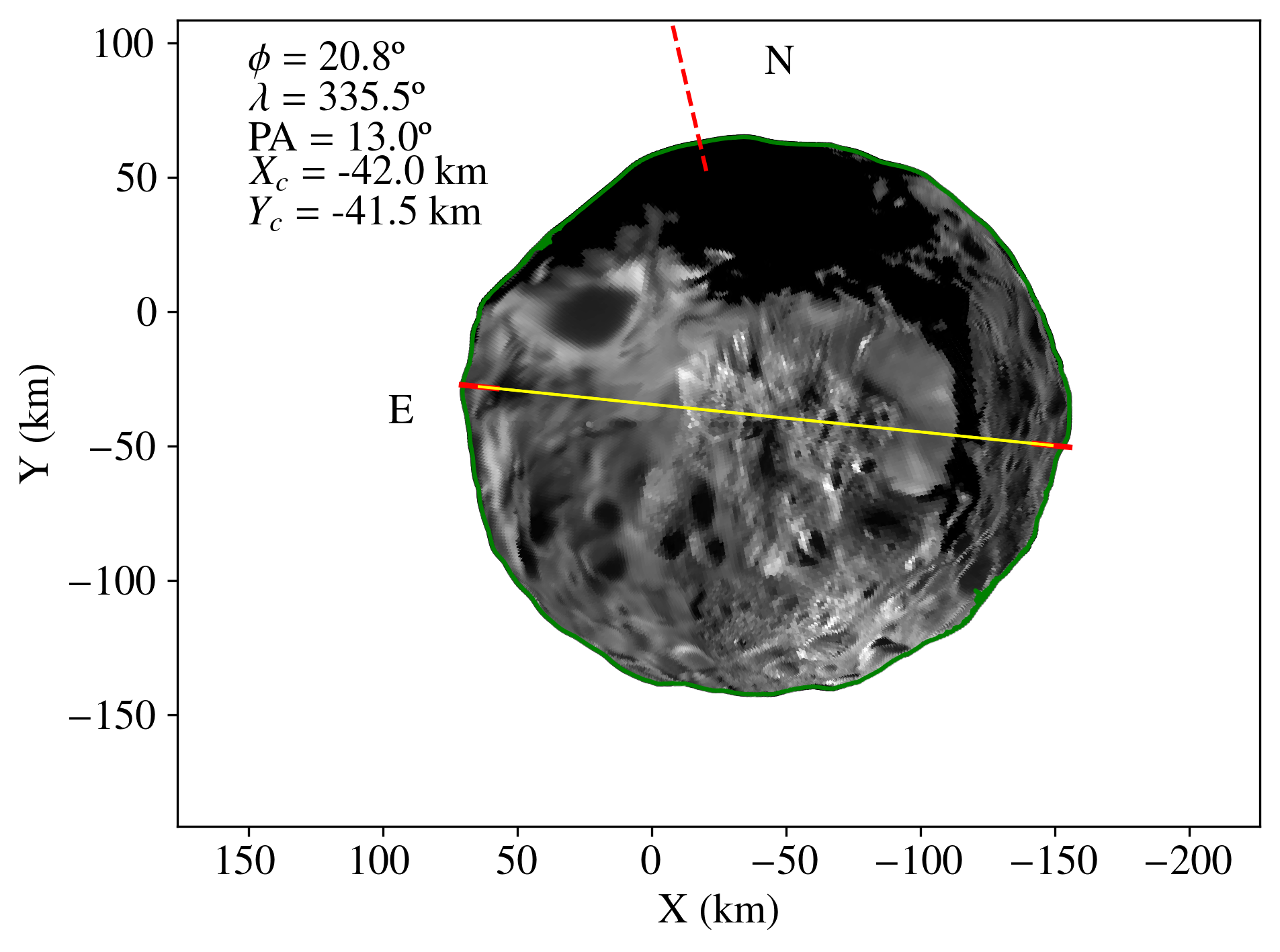}}
\caption{Similar to \autoref{Fig:Occ-Phoebe-jun19} for the 2018 August 13 occultation. For W2, the mean solution of the north and south solutions combined were considered.}
\label{Fig:Occ-Phoebe-aug13}
\end{figure*}

\begin{figure*}
\centering
\subfigure[][W1 solution]{\includegraphics[width=0.48\textwidth]{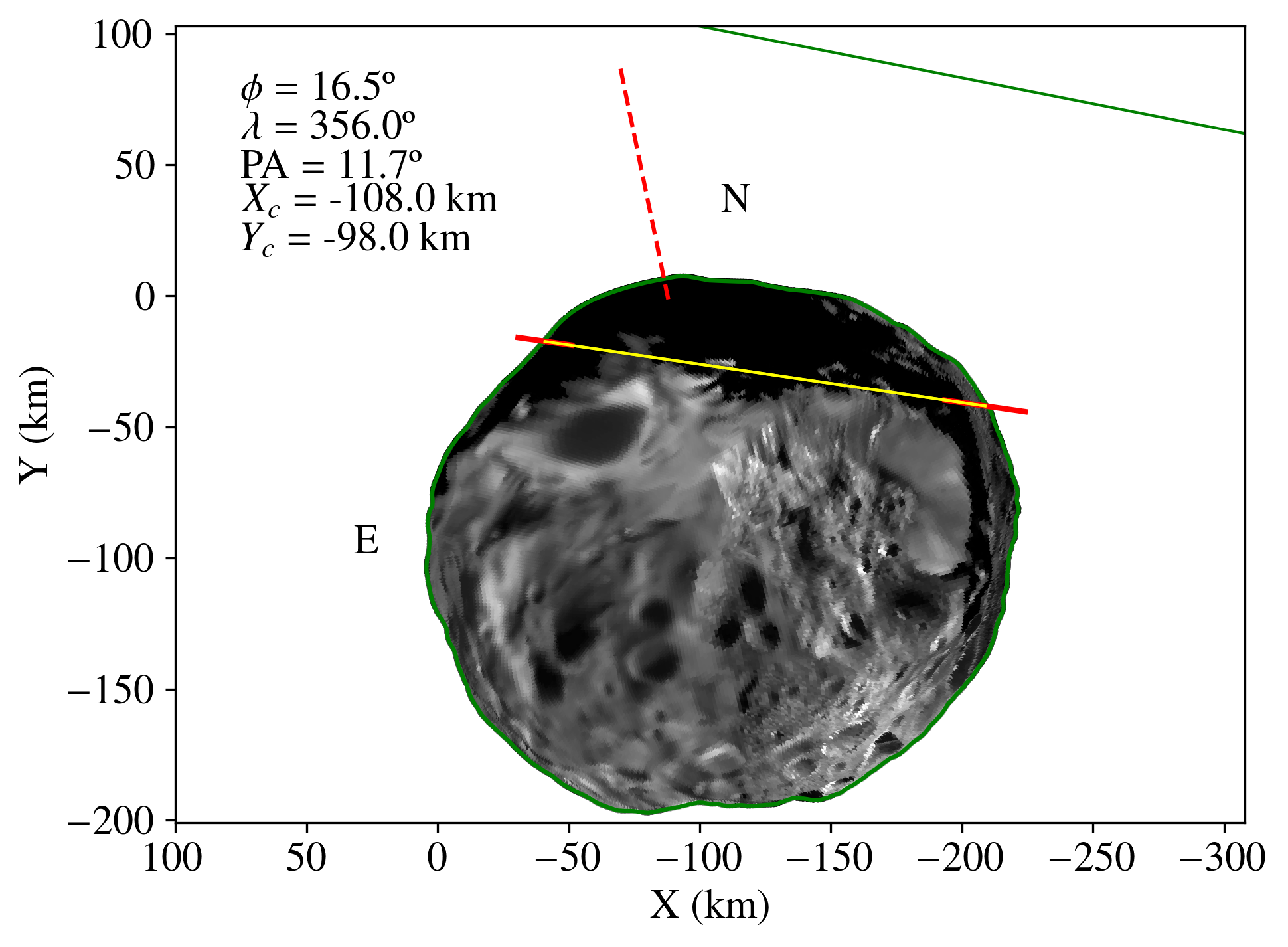}}
\subfigure[][W2 solution]{\includegraphics[width=0.48\textwidth]{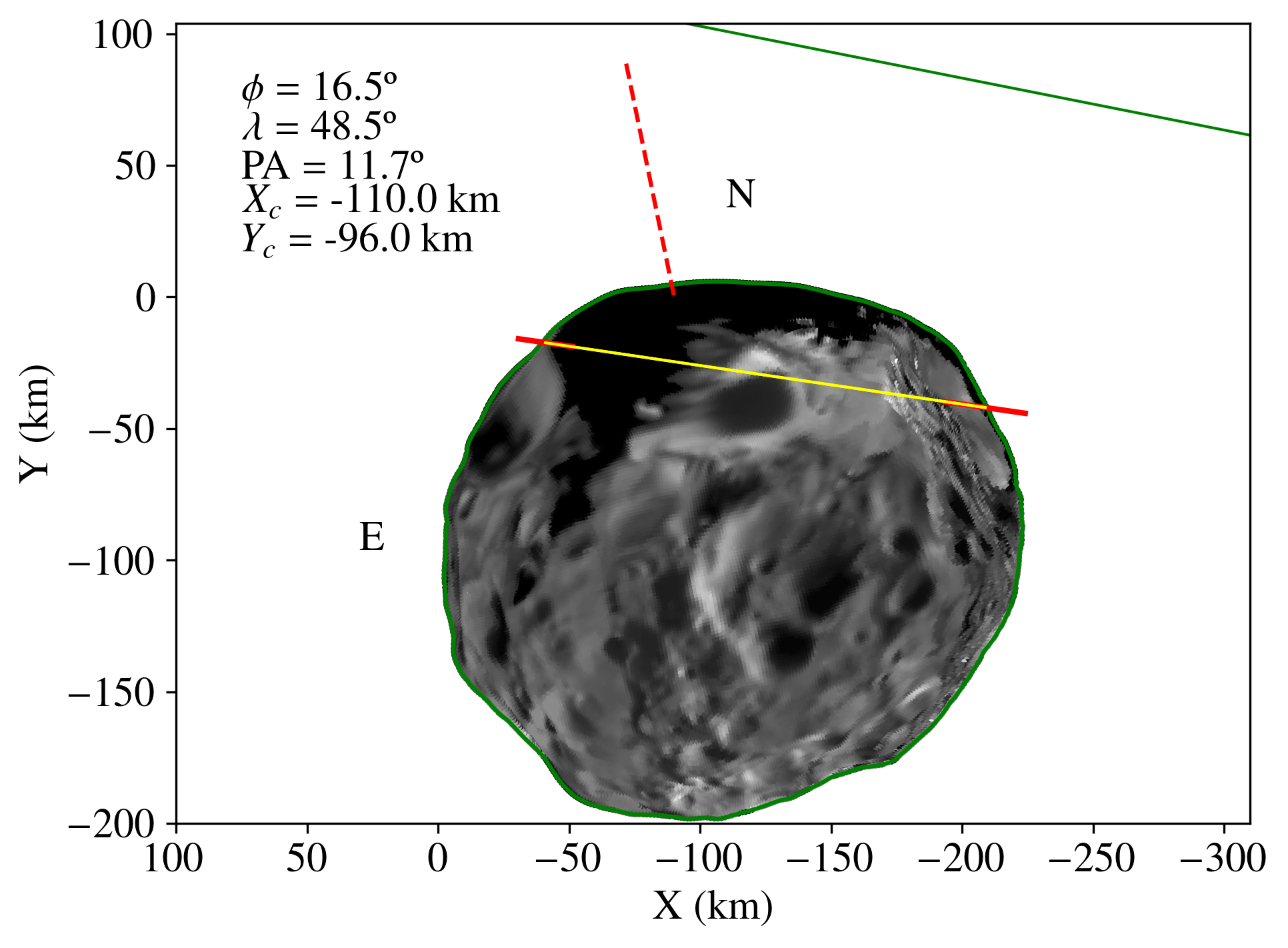}}
\caption{Similar to \autoref{Fig:Occ-Phoebe-jun19} for the 2019 June 07 occultation.}
\label{Fig:Occ-Phoebe-2019jun07}
\end{figure*}

For the occultation of 2019 June 07 the chord of Casleo would fit a north and south solution, however, the negative chord of SOAR eliminates the possibility of the south one. We also notice that the ephemeris offset has drifted from 2018. The best fits for these five occultations are presented in Figures \ref{Fig:Occ-Phoebe-jun19}-\ref{Fig:Occ-Phoebe-2019jun07}.

\section{Rotational Light Curve}
\label{Sec:Phoebe-light_curve}

To check the identified variation in the rotational phase, we obtained the rotational light curve of Phoebe in one night in 2017 from the 1.6m aperture f/10 Perkin-Elmer telescope and two consecutive nights in 2018 from the 0.6m aperture f/13.5 Boller \& Chivens telescope. Both telescopes are located at the Observatório do Pico dos Dias (OPD, 45\degr 34\arcmin 57\farcs5 W, 22\degr 32\arcmin 07\farcs8 S, 1864 m) run by Laboratório Nacional de Astrofísica/MCTI, Itajubá/MG, Brazil, IAU code 874. In 2017 July 28, we acquired about 5.5 hours of observations with an \textit{I} filter, while in 2018 May 23 and 24 we obtained about 5.5 and 7 hours respectively, both without the use of filters. All the observations were made with an IKON camera.

\begin{figure}
\begin{centering}
\includegraphics[width=\linewidth]{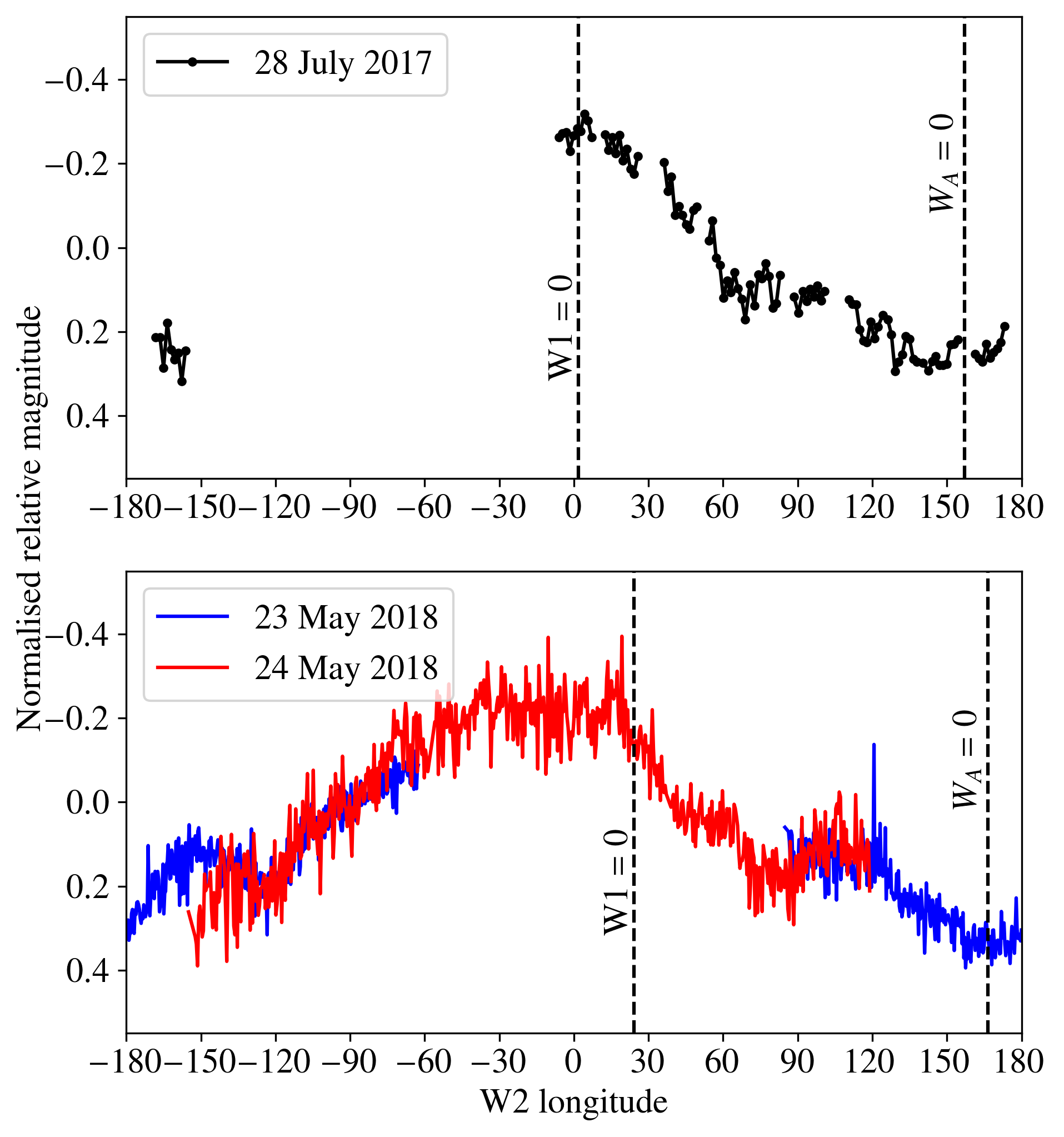}
\caption{Rotational light curves obtained from one night of observation in 2017 and two nights of observations in 2018 at the OPD. The abscissa is the sub-observer longitude using the W2 solution, as explained in the text. The location of the longitude $0\degr$ using the W1 solution and from Archinal et al ($W_A$) are also shown for comparison.}
\label{Fig:Phoebe_light_curve}
\end{centering}
\end{figure}

Because Phoebe is crossing a very dense region of stars, a Difference Image Analysis procedure was applied to the observation to remove the background stars\footnote{DIAPL Website: \url{https://users.camk.edu.pl/pych/DIAPL/index.html}}. The light curves were then obtained from the subtracted images using \textsc{praia}. \autoref{Fig:Phoebe_light_curve} shows the light curve (normalised \textcolor{red}{relative} magnitude) of three nights as a function of the sub-observer longitude obtained from the W2 solution. No fit was done to obtain a new period. The location of the longitude $0\degr$ using the W1 solution and from \cite{Archinal2018} ($W_A$) are also shown.

It is important to note that \cite{Bauer2004} assumed the maximum of the curve (maximum brightness) as the $0\degr$ longitude following the longitude system of \cite{Colvin1989}. We can see that the maximum in \autoref{Fig:Phoebe_light_curve} is closer to the $0\degr$ longitude for the W2 solution, while W1 and $W_A$ solutions are not.

\section{Results}
\label{Sec:Phoebe-results}

\subsection{Preferred Rotational Period}
\label{Subsec:Phoebe-preferred-rotation}

As shown in \autoref{Sec:Phoebe-rotation}, we managed to identify two possible solutions for the rotation period of Phoebe from the differences in sub-observer longitude determined from the 2017 occultation and the prediction using the rotation model of \cite{Archinal2018}. They are S1 = 9.274404 h $\pm$ 0.07 s and S2 = 9.273653 h $\pm$ 0.07 s, respectively related to the W1 and W2 solutions found in \autoref{Sec:Phoebe-rotation}.

In \autoref{Sec:Phoebe-2018}, we projected these values for the 2018 and 2019 occultations. The four events of 2018 were observed in an interval of one month, so they are expected to have similar ephemeris offsets. However, using the S1 rotational period solution, the derived position for the centre of Phoebe for the June 26 occultation is fairly inconsistent with the centres from the other occultations. The longitude of the June 19 event would also be very close to a region where the chord is larger than the apparent shape of Phoebe inside the 1-sigma error bar.

Also, as shown in \autoref{Sec:Phoebe-light_curve}, the maximum of the observed rotational light curve is also near the zero longitude using the S2 solution.

Given these facts and that the S2 solution is also closer to that published by \cite{Bauer2004}, S2 (with W2) is our preferred solution. Thus, the \textcolor{red}{final} rotational period found for Phoebe is \textcolor{red}{$9.27365 \pm 0.00002$ h}.

\subsection{Astrometric Results}
\label{Subsec:Phoebe-astrometry}

The fit of the chords to the shape model shows that the prediction of these events was accurate. The possibility to compare the chords with a shape model and the use of Gaia-DR2 stars allows for the determination of very precise astrometric positions for the satellite, with errors in the order of 1 mas.

\autoref{Tab:astrometry} shows the geocentric ICRS coordinates in the Gaia DR2 frame and respective uncertainties for Phoebe from the occultations. The associated sub-observed longitude is also shown in the equation for the preferred W2 (S2) solutions.

\begin{table*}
\caption{Geocentric ICRS coordinates in the Gaia DR2 frame, and respective uncertainties for Phoebe from the occultations as described in \autoref{Subsec:Phoebe-astrometry}. 
\label{Tab:astrometry}}
\begin{centering}
\begin{tabular}{lccc}
\hline  \hline
Date (UTC)  & $\alpha$ & $\delta$ &  $\lambda$ (W2) \tabularnewline
\hline
 2017 July 06 16:04:00 & $17^{\text{h}} ~31^{\text{m}} ~03\fs07802 \pm 0.3 \text{~mas}$ & $-22\degr ~00\arcmin ~57\farcs3585 \pm 0.4 \text{~mas}$ & $126.5\degr \pm 3.5\degr$\tabularnewline
 2018 June 19 04:38:00 & $18^{\text{h}} ~26^{\text{m}} ~16\fs38857 \pm 0.6 \text{~mas}$ & $-22\degr ~24\arcmin ~00\farcs5778 \pm 0.7 \text{~mas}$ & $251.5\degr \pm 3.7\degr$\tabularnewline
 2018 June 26 18:31:00 & $18^{\text{h}} ~24^{\text{m}} ~01\fs55924 \pm 1.1 \text{~mas}$ & $-22\degr ~26\arcmin ~12\farcs5432 \pm 0.4 \text{~mas}$ & $112.5\degr \pm 3.7\degr$\tabularnewline
 2018 July 03 13:38:00 & $18^{\text{h}} ~22^{\text{m}} ~00\fs25759 \pm 1.2 \text{~mas}$ & $-22\degr ~28\arcmin ~10\farcs7045 \pm 0.9 \text{~mas}$ & $325.0\degr \pm 3.7\degr$\tabularnewline
 2018 Aug 13 12:53:00 & $18^{\text{h}} ~12^{\text{m}} ~22\fs39941 \pm 1.9 \text{~mas}$ & $-22\degr ~38\arcmin ~44\farcs4513 \pm 6.7 \text{~mas}$ & $335.5\degr \pm 3.7\degr$\tabularnewline
 2019 June 07 03:54:00 & $19^{\text{h}} ~21^{\text{m}} ~18\fs59540 \pm 2.5 \text{~mas}$ & $-21\degr ~44\arcmin ~25\farcs6461 \pm 2.7 \text{~mas}$ & $48.5\degr \pm 4.0\degr$\tabularnewline
\hline
\end{tabular}
\par
\end{centering}
\end{table*}

\section{Conclusions}
\label{Sec:Phoebe-discussao}

The stellar occultation of 2017 July 06 was the first occultation by an irregular satellite ever observed. The observation was possible due to the improved ephemeris developed by \cite{GomesJunior2016} and the fortuitous passage of Phoebe in front of the galactic plane, which allowed to predict more events with bright stars.

Six stellar occultations were observed, one with two positive detections and other five single-chord events. The uncertainties obtained in the ingress and egress instants are of the order of 1 km and reflects the precision of the technique.

Due to the \textit{Cassini}'s observation of Phoebe, that provided a good knowledge of the shape of the satellite and showed its cratered nature, fitting circular or elliptical shapes to the chords would not provide accurate results. The 3D shape model of Phoebe from \cite{Gaskell2013} allowed a more precise analysis. Comparing the model with the occultation chords, we found that the chords of 2017 July 06 occultation better fit to the model when projecting the model in the sub-observer longitude $\lambda = 126.5\pm3.5\degr$.

We managed to determine \textcolor{red}{the} rotation period for Phoebe \textcolor{red}{from stellar occultations}. For that, we primarily used the differences in sub-observer longitude determined from the two-chord 2017 occultation and from the prediction using the rotation model of \cite{Archinal2018}. We eliminated the ambiguity in the solution by using the other single-chord occultations and rotation light curves observed by us and from  \cite{Bauer2004}. The \textcolor{red}{final} rotation period found is p = \textcolor{red}{$9.27365 \pm 0.00002$ h}. \textcolor{red}{Although the value is within the error bar of \cite{Bauer2004}, our results improve the determination of the rotational period of the satellite by one order of magnitude}. The related location of the prime meridian (the half great-circle connecting the body’s north and south pole defined as $\lambda=0\degr$ in the body-centred reference frame) is $W(\degr) = 125.56 + 931.6717d$, following the formalism by \cite{Archinal2018}.

We also obtained six geocentric ICRS positions as realised by the GAIA DR2 frame for Phoebe with kilometre accuracy. For comparison, the accuracy of \cite{Desmars2013} ephemeris during the \textit{Cassini} flyby is in the order of 10 km. The positions obtained here can be helpful to improve the ephemeris and, consequently, the occultation predictions.

The observation of more multi-chord stellar occultations by Phoebe can help to constrain its rotation period better, derive very precise positions and probe uncharted surface features in its northern hemisphere not mapped by \textit{Cassini}.  As predicted by \cite{GomesJunior2016}, the density of stellar occultations by Phoebe will decrease after 2019, since Saturn is leaving the apparent galactic plane. Even though the number of favourable events will decrease, we will have better ephemeris and will be able to focus on selected events.

These results with Phoebe give us more expertise and expectation for the campaign of stellar occultations by the Jovian irregular satellites in 2019-2020 when Jupiter, in turn, will be crossing the apparent plane of the Galaxy.

\section*{Acknowledgements}

A.R.G.J. thanks the financial support of CAPES, CNPq, INCT and FAPESP (proc. 2018/11239-8).
M.A. thanks CNPq (Grants 427700/2018-3, 310683/2017-3 and 473002/2013-2) and FAPERJ (Grant E-26/111.488/2013).
FBR acknowledges CNPq grant 309578/2017-5. G.B.R. is thankful for the support of CAPES/Brazil and FAPERJ (Grant E26/203.173/2016).
BM thanks the CAPES/Cofecub-394/2016-05 grant.
RV-M acknowledges grants: CNPq: 304544/2017-5, 401903/2016-8, Faperj: PAPDRJ-45/2013 and E-26/203.026/2015, Capes/Cofecub: 2506/2015.
J.I.B.C. acknowledges CNPq grant 308150/2016-3.
Based on observations obtained at the Southern Astrophysical Research (SOAR) telescope, which is a joint project of the Minist\'{e}rio da Ci\^{e}ncia, Tecnologia, Inova\c{c}\~{o}es e Comunica\c{c}\~{o}es (MCTIC) do Brasil, the U.S. National Optical Astronomy Observatory (NOAO), the University of North Carolina at Chapel Hill (UNC), and Michigan State University (MSU).
Based on data obtained at Complejo Astron\'omico El Leoncito, operated under agreement between the Consejo Nacional de Investigaciones Cient\'ficas y T\'ecnicas de la Rep\'ublica Argentina and the National Universities of La Plata, C\'ordoba and San Juan.
The work was based on observations made at the Laboratório Nacional de Astrofísica (LNA), Itajubá-MG, Brazil.
TRAPPIST-South is funded by the Belgian Fund for Scientific Research under the grant FRFC 2.5.594.09.F. EJ is a Belgian FNRS Senior Research Associates.
The work was based on observations made by MiNDSTEp from the Danish 1.54m telescope at ESO's La Silla observatory.
This collaboration, as part of the Encelade working group (http://www.issibern.ch/teams/encelade/), has been supported by the International Space Sciences Institute in Bern, Switzerland.
Part of the research leading to these results has received funding from the European Research Council under the European Communitys H2020 (2014-2020/ERC Grant Agreement no. 669416 “LUCKY STAR”).




\bibliographystyle{mnras}
\bibliography{Referencias} 








\bsp	
\label{lastpage}
\end{document}